\documentclass{aa}

\graphicspath{{./}{plots/}}

\usepackage[normalem]{ulem}
\usepackage{soul}
\usepackage{natbib,twoopt, xcolor, placeins}
\usepackage[breaklinks=true]{hyperref} 
\bibpunct{(}{)}{;}{a}{}{,}             
\makeatletter
  \newcommandtwoopt{\citeads}[3][][]{\href{http://adsabs.harvard.edu/abs/#3}%
    {\def\hyper@linkstart##1##2{}%
     \let\hyper@linkend\@empty\citealp[#1][#2]{#3}}}
  \newcommandtwoopt{\citepads}[3][][]{\href{http://adsabs.harvard.edu/abs/#3}%
    {\def\hyper@linkstart##1##2{}%
     \let\hyper@linkend\@empty\citep[#1][#2]{#3}}}
  \newcommandtwoopt{\citetads}[3][][]{\href{http://adsabs.harvard.edu/abs/#3}%
    {\def\hyper@linkstart##1##2{}%
     \let\hyper@linkend\@empty\citet[#1][#2]{#3}}}
  \newcommandtwoopt{\citeyearads}[3][][]%
    {\href{http://adsabs.harvard.edu/abs/#3}
    {\def\hyper@linkstart##1##2{}%
     \let\hyper@linkend\@empty\citeyear[#1][#2]{#3}}}
\makeatother

\newcommand{\edits}[1]{{#1}}

\begin{document}

\title{Empirical derivation of the metallicity evolution with time and radius using TNG50 Milky Way and Andromeda analogues}

\author{B. Ratcliffe\inst{1} 
\and S. Khoperskov\inst{1}
\and I. Minchev\inst{1}
\and L. Lu\inst{2}
\and R. S. de Jong\inst{1}
\and M. Steinmetz\inst{1}}
    
\institute{Leibniz-Institut für Astrophysik Potsdam (AIP), An der Sternwarte 16, 14482 Potsdam, Germany
\and American Museum of Natural History, Central Park West, Manhattan, NY, USA}

\date{Received \today; accepted ...}

\abstract
{Recent works use a linear birth metallicity gradient to estimate the evolution of the [Fe/H] profile in the Galactic disk over time, and infer stellar birth radii (R$_\text{birth}$) from [Fe/H] and age measurements. These estimates rely on the evolution of [Fe/H] at the Galactic center ([Fe/H](0, $\tau$)) and the birth metallicity gradient ($\nabla$[Fe/H]($\tau)$) over time --- quantities that are unknown and inferred under key assumptions.}
{In this work, we use the sample of Milky Way and Andromeda analogues from the TNG50 simulation to investigate the ability to recover [Fe/H](R, $\tau$) in a variety of galaxies.}
{Using stellar disk particles, we tested the assumptions required in estimating R$_\text{birth}$, [Fe/H](0, $\tau$), and $\nabla$[Fe/H]($\tau)$ using recently proposed methods to understand when they are valid.}
{We show that $\nabla$[Fe/H]($\tau)$ can be recovered in most galaxies to within 26\% from the range in [Fe/H] across age, with better accuracy for more massive and stronger barred galaxies. We also find that the true central metallicity is unrepresentative of the genuine disk [Fe/H] profile; thus we propose to use a projected central metallicity instead. About half of the galaxies in our sample do not have a continuously enriching projected central metallicity, with a dilution in [Fe/H] correlating with mergers. Most importantly, galaxy-specific [Fe/H](R, $\tau$) can be constrained and confirmed by requiring the R$_\text{birth}$ distributions of mono-age, solar neighborhood populations to follow inside-out formation.}
{We conclude that examining trends with R$_\text{birth}$ is valid for the Milky Way disk and similarly structured galaxies, where we expect R$_\text{birth}$ can be recovered to within 20\% assuming today's measurement uncertainties in TNG50.}

\authorrunning{Ratcliffe et al.}
\titlerunning{Recovering the disk metallicity evolution with TNG50}
\maketitle

\section{Introduction}
\label{sec:intro}

\nolinenumbers

The ultimate goal of Galactic archaeology --- the study of the history of star formation as recorded in stellar properties --- is to understand the formation and evolution of the Milky Way. To do so, though, we have to interpret its history from a present-day snapshot, which includes stellar kinematics, ages, and chemical abundances. Knowledge of the stars' time and place of birth would allow us to directly trace the evolution of the Galaxy over time.

It is well established that stars radially migrate away from their birth sites \citep[e.g.][]{Selwood2002, 2009schonrichBinney, Minchev2010, Frankel2018, Carr2022}, meaning that a given location in the Galaxy can be comprised of stars born at a variety of different radii depending on stellar age and migration rate (\citealt{Minchev2018_rbirth, Agertz2021_vintergatanI, Carrillo2023}). It has also been shown that radial migration affects stars relatively early in their lifetime \citep{Kubryk2013, Grand2016_radialMigration, 2020Feltzing}, and therefore this mixing of stars from different birth radii (\mbox{$\rm \text{R}_\text{birth}$}) at a given current location makes it challenging to unravel evolutionary events; radial abundance gradients become weaker \citep{Pilkington2012, Minchev2012, Minchev2013, Kubryk2013, Ratcliffe2023_enrichment} and chemo-kinematic relations can become blurred (e.g., \citealt{Minchev2014}). 

Due to gravitational interactions with spiral arms, the central bar, and infalling satellites, stellar angular momentum can be permanently changed, making it impossible to trace a star's position back to its birth site by simply using its current kinematics~\citep{Selwood2002, Minchev2006, Roskar2008_migration, Quillen2009, Minchev2010, Minchev2012a, DiMatteo2013,Khoperskov_bar_migration2018}. Chemical abundances, on the other hand, are an archaeological record of the physical conditions in the interstellar medium at the birth time and position of stars as they are relatively unchanging with time \citep{Tinsley1979}. The idea of strong chemical tagging utilizes these observable abundances to recover dispersed star clusters, which are believed to be chemically homogeneous \citep{2002freeman-BH, BH2010}. While recent works have shown that strong chemical tagging is infeasible with current-day observational uncertainties and sample size \citep{2013Lindegren,Ting2015, Casamiquela2021}, a weaker version of chemical tagging to recover stars born at similar times and places in the Galactic disk is more likely to succeed \citep{2020Ratcliffe, 2022Ratcliffe}.

Recent works leverage chemical abundances of disk stars to estimate the effect of radial migration in the Milky Way disk (e.g. \citealt{Minchev2018_rbirth, Frankel2018, 2019Frankel, Frankel2020, Lu2022_Rb, Ratcliffe2023_enrichment, Ratcliffe2023_chemicalclocks, Wang2023}). One way to understand this effect and recover the Milky Way's evolution is by estimating stellar birth radii from a star's metallicity and age. These \mbox{$\rm \text{R}_\text{birth}$} methods rely on a linear relationship between [Fe/H] and \mbox{$\rm \text{R}_\text{birth}$} at a given look-back time, where the slope (birth metallicity gradient over cosmic time; \mbox{$\rm \nabla [Fe/H](\tau)$}) and intercept (metallicity at a specific radius R over time; [Fe/H](R, $\tau$)) are unknown. Through the development of a largely model independent approach, \cite{Minchev2018_rbirth} simultaneously recovers stellar \mbox{$\rm \text{R}_\text{birth}$}, \mbox{$\rm \nabla [Fe/H](\tau)$}, and [Fe/H](R=R$_\odot$, $\tau$) by requiring that \mbox{$\rm \text{R}_\text{birth}$} distributions of mono-age populations remain non-negative and follow expectations of inside-out disk formation \citep{Matteucci1989}. With their discovery that the scatter in metallicity across age bins is linearly correlated with \mbox{$\rm \nabla [Fe/H](\tau)$}, \cite{Lu2022_Rb} allow for the determination of \mbox{$\rm \text{R}_\text{birth}$} to be predominantly self-contained, with the largest modeling requirement being that the evolution of [Fe/H] in the Galactic center smoothly and continuously enriched over time. 

In recovering \mbox{$\rm \nabla [Fe/H](\tau)$}, \cite{Lu2022_Rb} finds that the radial [Fe/H] gradient had a steepening event $8-11$ Gyr ago, which they attribute to the Gaia-Sausage-Enceladus (GSE; \citealt{Belokurov2018, Helmi2018_gse, Haywood2018}. See \citealt{Prantzos2023} for an alternative argument). Applying the same method to higher spectral resolution data, \cite{Ratcliffe2023_enrichment} find two additional steepening events about 4 and 6 Gyrs ago, which coincided with passages of the Sagittarius dwarf galaxy \citep{Ibata1994, Law2010}. The effect of mergers on the radial metallicity gradient is investigated using cosmological hydrodynamical simulations in \cite{Buck2023}, where it is found mergers brought in pristine gas and lowered the metallicity in the outer disk, therefore steepening the gradient (though potentially not over long time scales; \citealt{Renaud2024}). It is also shown that if the merger was large enough, the metallicity at the galactic center would decrease, in direct conflict with the monotonic [Fe/H] enrichment at the Galactic center used in Milky Way disk \mbox{$\rm \text{R}_\text{birth}$} derivations (e.g. \citealt{2020_buckchemical, Renaud2024}). 

Modern hydrodynamical cosmological simulations --- such as NIHAO-UHD \citep{2020buck_NIHAO-UHD}, Feedback In Realistic Environments-2 (FIRE-2; \citealt{2018FIRE2}), VINTERGATAN \citep{Agertz2021_vintergatanI, Renaud2021_vintergatanII, Renaud2021_vintergatanIII}, and HESTIA \citep{Libeskind2020, Khoperskov2023_Hestia} --- contain a wealth of information for investigating the effect of different evolutionary histories. In particular, TNG50 \citep{Pillepich2019, Nelson2019a_TNG, Nelson2019} contains a sample of 198 Milky Way and Andromeda-like galaxies that have diverse galactic properties and histories \citep{Pillepich2023_MW_Sample}. With access to a full set of stellar properties --- including those that are not measurable for Milky Way stars (e.g. \mbox{$\rm \text{R}_\text{birth}$}) --- for a large variety of simulated Milky Way-like galaxies, we can investigate the dependencies of different evolutionary histories on our ability to recover stellar birth radii. 

Previous works have only tested the ability to derive birth radii on a handful of simulated galaxies, with little to no investigation into how the central metallicity evolves with time. In this work, we further tested the assumptions required to estimate birth radii using the large Milky Way and Andromeda analogue sample from TNG50. Thanks to the large quantity of galaxies in this sample, we were able to test the validity of the assumptions on a diverse set of formation and evolutionary histories and explore which conditions are potentially violated in the Milky Way. This paper is organized as follows. Discussions of the data and \mbox{$\rm \text{R}_\text{birth}$} method used for this work are given in Sections \ref{sec:data} and \ref{sec:method}. Sections \ref{sec:results_gradFeh} and \ref{sec:results_feh0} present our findings on the evolution of the metallicity gradient and central metallicity assumptions, respectively. The discussion of our work in context with the Milky Way is given in Section \ref{sec:discussion}. Section \ref{sec:conclusion} gives our conclusions.

\begin{figure*}
     \includegraphics[width=17cm]{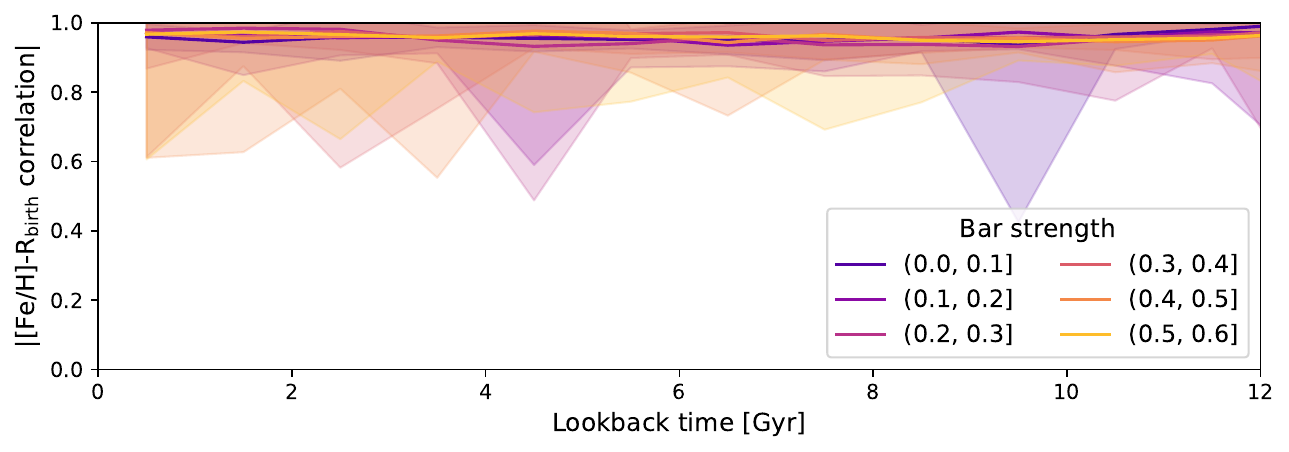}
\caption{Median absolute correlation between [Fe/H] and \mbox{$\rm \text{R}_\text{birth}$} for galaxies in the TNG50 Milky Way and Andromeda catalog across lookback time grouped by bar strength, with the shaded areas showing $\pm 1$ standard deviation about the median. The correlation between [Fe/H] and \mbox{$\rm \text{R}_\text{birth}$} was calculated for age bins of 1 Gyr. To accommodate for outliers and higher feedback, the correlation was calculated for the running median, using \mbox{$\rm \text{R}_\text{birth}$} bins of 0.25 kpc. We do not find that bar strength has an impact on the linearity of the average birth metallicity gradient. The large variability in the correlation between [Fe/H] and \mbox{$\rm \text{R}_\text{birth}$} for stronger barred galaxies at smaller lookback times is predominantly due to weak metallicity gradients, even though the gradient itself is linear.}
\label{fig:corr_feh_rb}
\end{figure*}

\section{Milky Way and Andromeda analogues from TNG50} \label{sec:data}

The TNG50 cosmological simulation \citep{Pillepich2019, Nelson2019a_TNG, Nelson2019} is the highest resolution run of the IllustrisTNG project \citep{Pillepich2018_TNG, Marinacci2018_TNG, Naiman2018_TNG, Springel2018_TNG, Nelson2018_TNG, Nelson2019a_TNG}, \edits{and was run with the moving-mesh code AREPO \citep{Springel2010}.} The simulation evolved gas, stars, dark matter, black holes, and magnetic fields inside a box of width 51.7 comoving Mpc from redshift $\approx$ 127 to redshift 0, with a resolution of $m_\text{baryon} = 8.5 \times 10^4 \text{M}_\odot$, $m_\text{DM} = 4.5 \times 10^5 \text{M}_\odot$, and a spatial resolution of the star-forming ISM gas of $\sim 100-140$ pc \citep{Pillepich2019}. \edits{Chemical enrichment of the interstellar medium accounted for heavy element nucleosynthesis from core-collapse supernovae (SNII), Type-Ia supernovae (SNIa), and asymptotic giant branch stars \citep{Pillepich2018_TNG, Naiman2018_TNG}. The yield tables were taken from \cite{Karakas2010, Doherty2014, Fishlock2014} (AGB), \cite{Portinari1998, Kobayashi2006} (SNII), and \cite{Nomoto1997} (SNIa) (see Table 2 in \citealt{Pillepich2018_TNG} for more information). The mixing processes were treated naturally in AREPO, with lower rates of numerical diffusion than fixed grid codes \citep{Springel2010, Naiman2018_TNG}.}

Within TNG50 there are 198 \edits{galaxies that are similar to the Milky Way or Andromeda, and} were selected using the following selection criteria at redshift 0 \edits{given in \cite{Pillepich2023_MW_Sample}}: 
\begin{itemize}
    \item The stellar mass of the galaxy is in the range $M_*(< 30$kpc) = $10^{10.5-11.2}$M$_\odot$;
    \item Disk-like morphology with spiral arms\edits{, chosen such that either the minor-to-major axis ratio of the galaxy’s stellar mass distribution is smaller than 0.45 or the galaxy appears disk-like and exhibits spiral arms through visual inspection of three-band images in edge-on and face-on projections};
    \item No other galaxy with stellar mass $\geq 10^{10.5}$M$_\odot$ is within 500kpc and the total mass of the halo host is smaller than $M_{200c}$(host)$<10^{13}$M$_\odot$.
\end{itemize}

We partner the Milky Way and Andromeda-like Satellites catalog\footnote{\edits{https://www.tng-project.org/data/milkyway+andromeda/}} \citep{Pillepich2023_MW_Sample} with bar strength measurements from \cite{Khoperskov2023} (taken as the peak value of $m$ = 2 Fourier harmonics of the density distribution) and disk scale lengths from \cite{SotilloRamos2022} (measured by fitting an exponential profile to the radial stellar surface density distribution excluding the bulge region) at redshift 0. Merger history information for each subhalo was recovered using SUBLINKGAL \citep{RodriguezGomez2015}. The gas mass ratio of a merger is defined as the ratio of the gas mass of the secondary progenitor to the first progenitor at the time when the secondary progenitor achieves its maximum gas mass\footnote{https://www.tng-project.org/data/forum/topic/662/nummerger-in-sublink/}.

To make the best comparisons of this sample with the Milky Way, we scaled each galaxy to have the same disk scale length (3.5 kpc; \citealt{BH2016}) and rotational velocity at the sun (238 km/s; \citealt{BH2016}) as the Milky Way disk. We also performed the following selection cuts to capture the stellar disk: $\mbox{$\rm \text{R}_\text{birth}$} < 15$ kpc, R $<15$ kpc, $|z| < 1$ kpc, |z$_\text{birth}| < 1$ kpc, ecc $<0.5$. Eccentricity was calculated using the following equation: 
$$\text{ecc} = \sqrt{v_r^2 + 2(v_\phi-v_0)^2}/(\sqrt{2}v_0),$$
where $v_r$ and $v_\phi$ are the radial and azimuthal velocities of the stellar particle, and $v_0$ is the rotational velocity of the galaxy at 2.2 disk scale lengths. We only used stellar particles that were born bound and are currently bound to the host galaxy to ensure stellar particles formed in situ. We also scaled the [Fe/H] abundance to the solar value from \cite{Asplund2009}. Due to uncertainties in chemical yields and modeling in simulations \citep{2021BuckHD_chemEnrich}, we focus our analysis on the relative values of [Fe/H] over time, and not the absolute values.

\section{Birth radii Method} \label{sec:method}

The goal of this paper is to extensively investigate the assumptions required to recover the evolution of the metallicity profile in galactic disks and additionally stellar birth radii. The recent method proposed by \cite{Lu2022_Rb} (based on \citealt{Minchev2018_rbirth}) assumes that for any lookback time ($\tau$) we can write [Fe/H] as a function of the metallicity gradient at that time (\mbox{$\rm \nabla [Fe/H](\tau)$}), birth radius, and the metallicity at the Galactic center (\mbox{$\rm [Fe/H](0, \tau$)}):
\begin{equation} \label{eqn:feh_Rb}
\mbox{$\rm [Fe/H]$}(\mbox{$\rm \text{R}_\text{birth}$}, \tau) = \mbox{$\rm \nabla [Fe/H](\tau)$} \mbox{$\rm \text{R}_\text{birth}$} + \mbox{$\rm [Fe/H](0, \tau$)}.
\end{equation}
Then, by rearranging Equation \ref{eqn:feh_Rb}, \mbox{$\rm \text{R}_\text{birth}$} can be found as a function of metallicity and age: 
\begin{equation} \label{eqn:rb}
\mbox{$\rm \text{R}_\text{birth}$}(\text{age}, \mbox{$\rm [Fe/H]$}) = \frac{\mbox{$\rm [Fe/H]$} - \mbox{$\rm [Fe/H](0, \tau$)}}{\mbox{$\rm \nabla [Fe/H](\tau)$}}.
\end{equation}
The main assumptions required can be broken down into two categories: assumptions regarding (1) the metallicity gradient and (2) the central metallicity. Our first results section (Section \ref{sec:results_gradFeh}) investigates the assumptions surrounding the metallicity gradient, while our second results section (Section \ref{sec:results_feh0}) explores the assumptions involved in the evolution of \mbox{$\rm [Fe/H](0, \tau$)}. Section \ref{sec:discussion_MWsuccess} puts this method in context with the Milky Way.

\section{Results I: recovering the evolution of the [Fe/H] radial gradient} \label{sec:results_gradFeh}

We start by examining the assumptions required to recover the evolution of the radial metallicity gradient with time. Namely, we tested the validity of a linear relationship between [Fe/H] and birth radius (Section \ref{sec:grad_Rbcorr}) and uncovered which galaxies show a correlation between the metallicity gradient and range in metallicity (\rm \mbox{$\text{Range}\widetilde{\mbox{$\rm [Fe/H]$}}(age)$}) as first proposed in \cite{Lu2022_Rb} (Section \ref{sec:grad_Scatcorr}).

\begin{figure*}
     \centering
     \includegraphics[width=.85\textwidth]{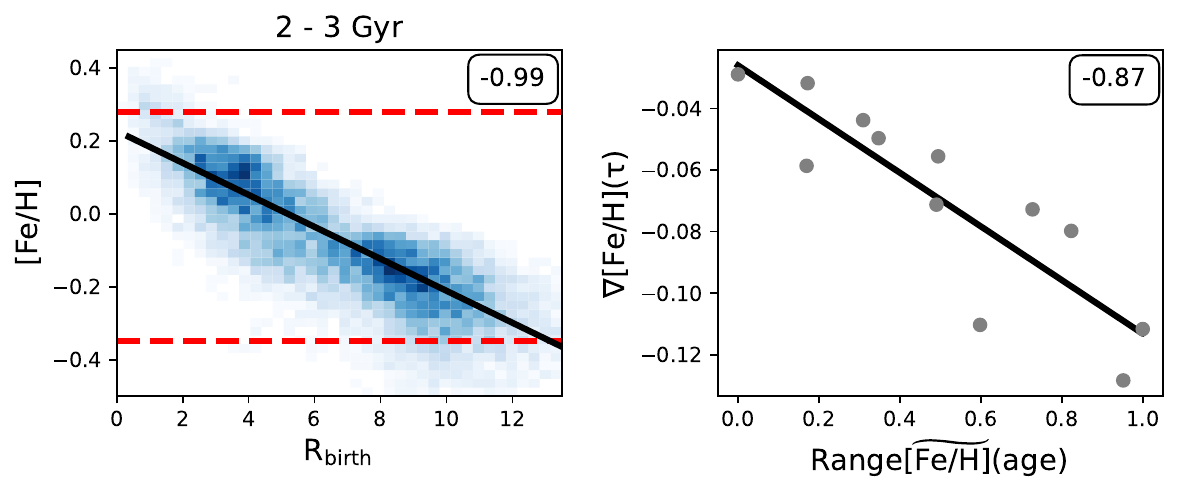}
\caption{Example (subhalo 577372) illustrating the recovery of \mbox{$\rm \nabla [Fe/H](\tau)$} from \rm \mbox{$\text{Range}\widetilde{\mbox{$\rm [Fe/H]$}}(age)$}. \textbf{Left}: Density distribution of the [Fe/H]--\mbox{$\rm \text{R}_\text{birth}$} plane for stars born $2-3$ Gyr ago. The correlation between the [Fe/H] running median and \mbox{$\rm \text{R}_\text{birth}$} is given in the top right corner. The fitted true birth gradient is shown as the black line, and the maximum and minimum [Fe/H] used to determine the range are shown as the dashed red lines. \textbf{Right}: \mbox{$\rm \nabla [Fe/H](\tau)$} versus \rm \mbox{$\text{Range}\widetilde{\mbox{$\rm [Fe/H]$}}(age)$} for the same galaxy, with each point representing a different age bin (width 1 Gyr). \rm \mbox{$\text{Range}\widetilde{\mbox{$\rm [Fe/H]$}}(age)$} is determined by looking at the range in [Fe/H] across all age bins, and then normalized so the minimum range is given a value of 0, and the maximum range is given a value of 1. The correlation between \rm \mbox{$\text{Range}\widetilde{\mbox{$\rm [Fe/H]$}}(age)$} and \mbox{$\rm \nabla [Fe/H](\tau)$} for this galaxy is -0.87, indicating the evolution of the gradient can be recovered well from \rm \mbox{$\text{Range}\widetilde{\mbox{$\rm [Fe/H]$}}(age)$}.}
\label{fig:rangeGrad_example}
\end{figure*}

\subsection{Linear relationship between [Fe/H] and birth radius}\label{sec:grad_Rbcorr}

A necessary requirement in the ability to recover stellar birth radii is a linear birth metallicity gradient throughout the galactic disk. This is supported in observations of young stars in the Milky Way \citep{Deharveng2000, Esteban2017, ArellanoCordova2021} and is shown in some simulations \citep{Vincenzo2018, Lu2022_sims}. However, some simulations exhibit a non-linear relationship between [Fe/H] and \mbox{$\rm \text{R}_\text{birth}$}, with the inner few kpc having a steeper slope \citep[e.g.][]{Agertz2021_vintergatanI, Bellardini2022}. Therefore, we begin our analysis by investigating the likelihood of a galaxy hosting a linear birth metallicity gradient in the TNG50 Milky Way and Andromeda analogues sample. 

Figure \ref{fig:corr_feh_rb} shows the median correlation (with the shaded region representing $\pm 1$ standard deviation) between [Fe/H] and \mbox{$\rm \text{R}_\text{birth}$} over lookback time colored by bar strength as defined in \cite{Khoperskov2023}. Since the purpose of this work is to investigate our ability of recovering birth radii with the method used in \cite{Lu2022_Rb, Ratcliffe2023_enrichment, Ratcliffe2023_chemicalclocks}, we chose to examine the correlation in age bins of 1 Gyr (similar trends are shown for age bins of 0.5 Gyr). To accommodate for outliers and strong migration during this period (Section \ref{sec:smallScatt}), we looked at the correlation of the [Fe/H] and \mbox{$\rm \text{R}_\text{birth}$} running median. \edits{Therefore, the correlation is defined as $$\frac{\sum_{i=1}^n(\text{R}_{\text{birth}, i} - \overline{\text{R}_\text{birth}})(\text{[Fe/H]}_i - \overline{\text{[Fe/H]}})}{\sum_{i=1}^n(\text{R}_{\text{birth}, i} - \overline{\text{R}_\text{birth}})^2\sum_{i=1}^n(\text{[Fe/H]}_i - \overline{\text{[Fe/H]}})^2},$$ where $n$ is the number of radial bins used of width 0.25 kpc, [Fe/H]$_i$ and $\text{R}_{\text{birth}, i}$ represent the median [Fe/H] and average $\text{R}_\text{birth}$ of the radial bin, and the bar denotes the average R$_\text{birth}$ and [Fe/H] of that bin.} Overall, we find a high correlation between [Fe/H] and \mbox{$\rm \text{R}_\text{birth}$} regardless of bar strength, in agreement with previous works on linear birth metallicity gradients (e.g. \citealt{Lu2022_sims}). However, unlike \cite{Lu2022_sims}, we find that a strong linear metallicity gradient exists already at early times when the photo-galactic disc is just starting to form.

While the median [Fe/H] gradient is linear across the TNG50 Milky Way and Andromeda-like galaxies of our sample, there are a number of galaxies which have a lower correlation between metallicity and birth radius as shown by the shaded regions in Figure \ref{fig:corr_feh_rb}. Many of these galaxies are given an artificially low correlation value due to their seemingly flat [Fe/H] gradients in the past few Gyr, even though their gradient is technically linear. Another key factor in our determination of linearity of the metallicity gradient is the number of disk particles. About 20 galaxies have few stellar disk particles with younger ages, with some having nearly no stellar particles with ages $<4$ Gyr. This may be related to late massive mergers destroying the disk, which is found in 18 TNG50 analogues \citep{SotilloRamos2022}.

Since a non-linear metallicity gradient prevents the recovery of birth radii, we only examined galaxies that have a correlation between \mbox{$\rm \text{R}_\text{birth}$} and the [Fe/H] running median $< -0.85$. This corresponds to times when $\sim 75\%$ of the variation in [Fe/H] is predictable from \mbox{$\rm \text{R}_\text{birth}$}. To accommodate for weak gradients that are given low correlation values despite being linear, we choose to ignore those years when deciding how many recent Gyrs the disk had a linear gradient. We, therefore, removed 17 galaxies (8.5\% of our sample) due to their weak correlation between [Fe/H] and \mbox{$\rm \text{R}_\text{birth}$} in the most recent Gyrs for the remainder of this section. These 17 galaxies span a variety of bar strengths, and appear to be non-Milky Way-like in a variety of ways (see Section \ref{sec:appendix_lowCor} in Appendix for more information).

\subsection{Correlation between the metallicity gradient and scatter} \label{sec:grad_Scatcorr}

\begin{figure}
    \centering
     \includegraphics[width=.495\textwidth]{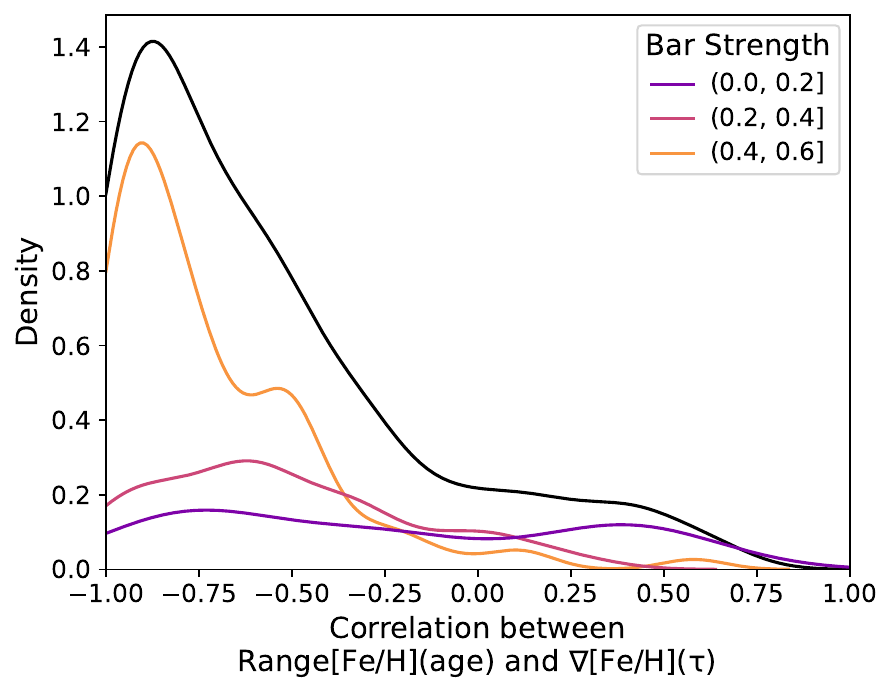}
\caption{Density distribution of the correlation between \rm \mbox{$\text{Range}\widetilde{\mbox{$\rm [Fe/H]$}}(age)$} and \mbox{$\rm \nabla [Fe/H](\tau)$} for 179 galaxies (black). These galaxies are chosen such that they have a linear gradient for (at least) their most recent 4 Gyr of star formation. Most galaxies exhibit a reasonable ability to have their \mbox{$\rm \nabla [Fe/H](\tau)$} evolution recovered from \rm \mbox{$\text{Range}\widetilde{\mbox{$\rm [Fe/H]$}}(age)$} (correlation $\sim -1$). This ability has a correlation with bar strength (colored lines), where stronger barred galaxies have a better recovery rate than weaker ones.}
\label{fig:scattGrad_hist}
\end{figure}

\begin{figure*}
    \centering
    \includegraphics[width=.425\textwidth]{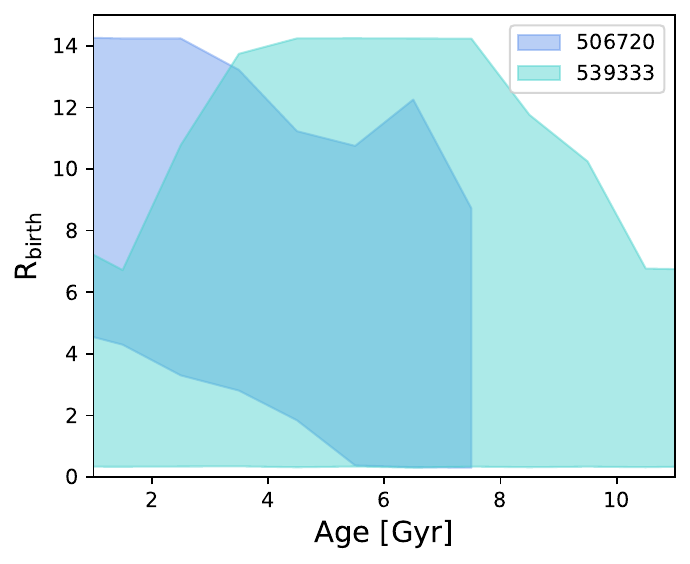}
     \includegraphics[width=.5\textwidth]{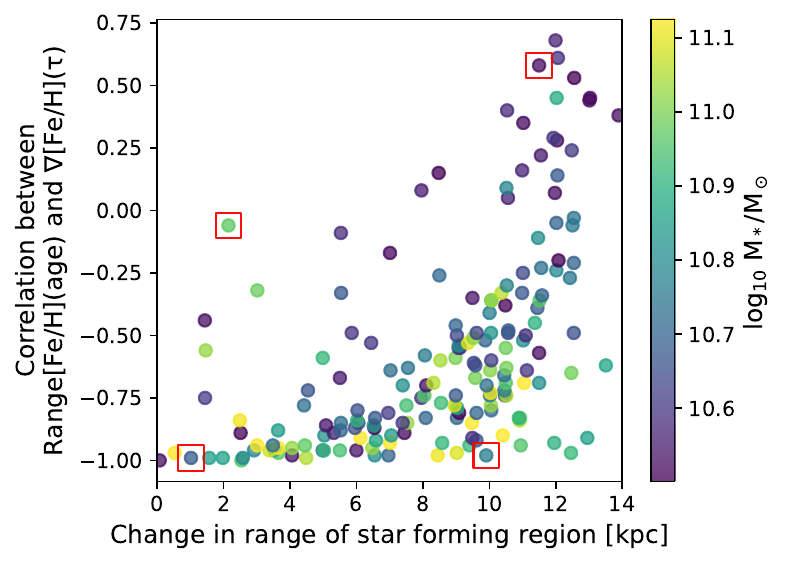}
\caption{\textbf{Left:} Star-forming region for 2 example galaxies (subhalos 506720 and 539333). The shaded region at a specific age indicates the radial range of stars forming during that time. The star-forming region of some galaxies (such as 539333) can vary in width of many kpc, while other galaxies (such as 506720) form stars in a much more consistent range. \textbf{Right:} Correlation between the scatter in [Fe/H] and the metallicity birth gradient as a function of how much the radial range of the star formation (RRSF) varies over time (defined as the difference between the maximum and minimum range stars are formed in the disk over time), colored by the galaxy's stellar mass within 30 kpc. A value of 0 indicates that the RRSF was constant over time (albeit potentially covering different radii such as subhalo 506720 in the left panel), while a larger value indicates that the galaxy formed stars over a few kpc some Gyrs, and over a much larger region other Gyrs (such as subhalo 539333 in the left panel). Overall, the more the range of the star-forming region varies over time, the less likely \mbox{$\rm \nabla [Fe/H](\tau)$} can be recovered from \rm \mbox{$\text{Range}\widetilde{\mbox{$\rm [Fe/H]$}}(age)$}. There is also a gradient with stellar mass, with higher mass galaxies showing a better ability to have their metallicity gradients recovered from \rm \mbox{$\text{Range}\widetilde{\mbox{$\rm [Fe/H]$}}(age)$}. The 4 highlighted galaxies in red boxes are examined in detail in Figure \ref{fig:outliers}.}
\label{fig:scattGrad_dffR}
\end{figure*}

\cite{Lu2022_Rb} shows that in the NIHAO-UHD \citep{2020buck_NIHAO-UHD} and Hestia simulations \citep{Libeskind2020, Khoperskov2023_Hestia}, there is a correlation between the scatter in [Fe/H] across different age bins and the evolution of the metallicity gradient with time. An example illustrating this correlation for a TNG50 galaxy is given in Figure \ref{fig:rangeGrad_example}. The density of the [Fe/H]--\mbox{$\rm \text{R}_\text{birth}$} plane for stars born $2-3$ Gyr ago for this example galaxy is depicted in the left panel, with the fitted gradient overlaid in black. The red dashed lines correspond to the boundary of the measured range in [Fe/H] during this age bin. The right panel shows \mbox{$\rm \nabla [Fe/H](\tau)$} vs \rm \mbox{$\text{Range}\widetilde{\mbox{$\rm [Fe/H]$}}(age)$} for the same galaxy, with each point corresponding to a different age bin. \rm \mbox{$\text{Range}\widetilde{\mbox{$\rm [Fe/H]$}}(age)$} here refers to the normalized Range in [Fe/H] over time, where the time of the largest range corresponds to \rm \mbox{$\text{Range}\widetilde{\mbox{$\rm [Fe/H]$}}(age)$} = 1 and the time of the smallest range has \rm \mbox{$\text{Range}\widetilde{\mbox{$\rm [Fe/H]$}}(age)$} = 0. 

The existence of the correlation between metallicity range and gradient allows for the ability to recover the evolution of the metallicity gradient with cosmic time directly from the data with minimal assumptions. This finding seems natural; the steeper the gradient, the larger the range in [Fe/H]. In our TNG50 sample of Milky Way and Andromeda-like galaxies, we find that most galaxies show a decent correlation between the metallicity gradient and [Fe/H] range (Figure \ref{fig:scattGrad_hist}), with a median correlation of -0.69. \edits{Here, a correlation of -1 implies that \mbox{$\rm \nabla [Fe/H](\tau)$} can be perfectly recovered from the scatter in [Fe/H] where the increase in scatter implies a steeper metallicity gradient (see the right panel of Figure 2 as an example of a correlation of -0.86), a correlation of 0 indicates that the scatter in [Fe/H] has no relation with the evolution of the metallicity gradient, and a correlation of 1 implies that the gradient steepens as the range decreases.} We also find that this result depends on bar strength, where the stronger barred galaxies have a better ability at recovering \mbox{$\rm \nabla [Fe/H](\tau)$} from \rm \mbox{$\text{Range}\widetilde{\mbox{$\rm [Fe/H]$}}(age)$} than weaker barred galaxies.

While most of the sample follows the natural assumption that the metallicity range and gradient are negatively correlated, there are a few galaxies that show little to no correlation between the two variables. There are even some galaxies that show the exact opposite of what is expected; that is, \rm \mbox{$\text{Range}\widetilde{\mbox{$\rm [Fe/H]$}}(age)$} and \mbox{$\rm \nabla [Fe/H](\tau)$} are positively correlated, indicating that a steeper gradient has a smaller range in [Fe/H]. These discrepancies are due to an underlying assumption in the ability to connect the scatter in [Fe/H] at a given age to the metallicity gradient at that lookback time --- the star-forming region of the disk must be the same radial range across cosmic time. This assumption is not necessarily true in our TNG50 sample (see left panel of Figure \ref{fig:scattGrad_dffR}), though it is crucial because the method claims that a steeper radial metallicity slope creates a larger spread in [Fe/H] than a weaker slope (i.e. a negative correlation). However, if the radial range of star formation (RRSF) was larger at the time of the weaker slope, the scatter in [Fe/H] could be larger. The right panel of Figure \ref{fig:scattGrad_dffR} illustrates the role that the difference in RRSF plays in our ability to connect \rm \mbox{$\text{Range}\widetilde{\mbox{$\rm [Fe/H]$}}(age)$} and \mbox{$\rm \nabla [Fe/H](\tau)$}. There is a trend that, on average, the ability to recover the metallicity gradient from the range in [Fe/H] decreases as the RRSF varies drastically over lookback time for a galaxy. There also appears to be a mass dependence, with more massive galaxies showing a better ability to recover \mbox{$\rm \nabla [Fe/H](\tau)$} from \rm \mbox{$\text{Range}\widetilde{\mbox{$\rm [Fe/H]$}}(age)$} (similar to what is seen in the NIHAO simulation; \citealt{Lu2024_RbLMC}). 

\begin{figure}
    \centering
     \includegraphics[width=.5\textwidth]{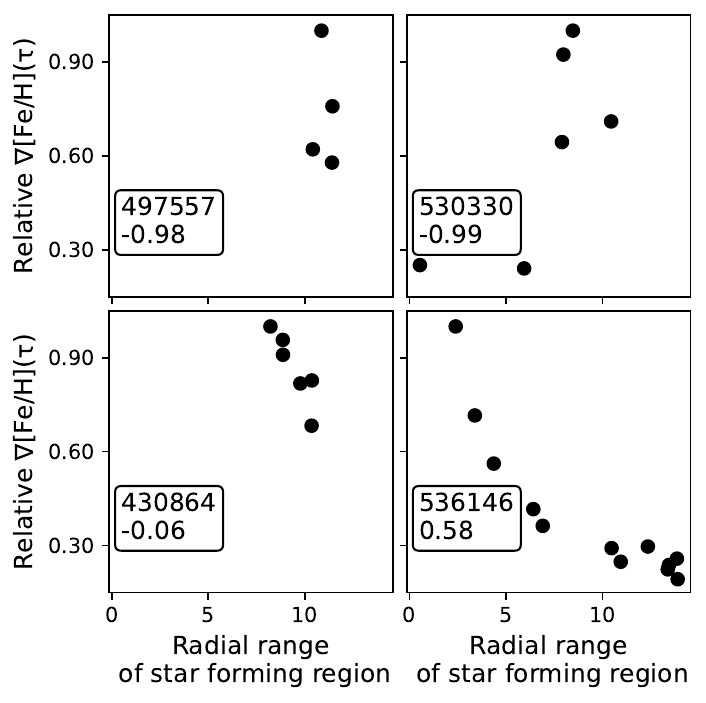}
\caption{Radial range of star-forming region (RRSF) versus the relative strength of the metallicity gradient for different ages, for the four highlighted galaxies in Figure \ref{fig:scattGrad_dffR}. The correlation between \mbox{$\rm \nabla [Fe/H](\tau)$} and \rm \mbox{$\text{Range}\widetilde{\mbox{$\rm [Fe/H]$}}(age)$} is given in the bottom left of each panel, below the subhalo ID. The top panels depict situations in which the evolution of the metallicity gradient can be recovered from the scatter in [Fe/H], while the bottom row illustrates two examples of galaxies where the gradient cannot be recovered in this fashion due to the gradient being steepest during a time of a lower star-forming region.}
\label{fig:outliers}
\end{figure}

To illustrate how the \rm \mbox{$\text{Range}\widetilde{\mbox{$\rm [Fe/H]$}}(age)$}--gradient correlation can be affected by a non-uniform RRSF over time, Figure \ref{fig:outliers} shows the relative strength of the [Fe/H] gradient versus the RRSF, with the correlation between \rm \mbox{$\text{Range}\widetilde{\mbox{$\rm [Fe/H]$}}(age)$} and the metallicity gradient shown in the bottom left corner. The four galaxies chosen are highlighted in red in the right panel of Figure \ref{fig:scattGrad_dffR}, and show examples of when the method works regardless of variation in RRSF over time (top row of Figure \ref{fig:outliers}) and when the method does not work (bottom row of Figure \ref{fig:outliers}). Specifically, the top left panel depicts a galaxy whose RRSF stays relatively constant with time. Thus, as the [Fe/H] gradient evolves, there is a direct correlation with the range in [Fe/H]. The top right panel also shows a galaxy with a strong ability to recover the [Fe/H] gradient from the range, but the RRSF increases by $\sim 8$ kpc throughout its lifetime. Here, the metallicity gradient is weakest when the RRSF is smallest, and the gradient steepens with increasing RRSF, further enforcing a strong correlation between \rm \mbox{$\text{Range}\widetilde{\mbox{$\rm [Fe/H]$}}(age)$} and \mbox{$\rm \nabla [Fe/H](\tau)$}.

There are some outliers in Figure \ref{fig:scattGrad_dffR} that show \mbox{$\rm \nabla [Fe/H](\tau)$} cannot be recovered using the method of \cite{Lu2022_Rb} even with a minimally varying RRSF. The bottom row of Figure \ref{fig:outliers} depicts two examples where the method does not work; one of a galaxy with essentially no relation between \mbox{$\rm \nabla [Fe/H](\tau)$} and \rm \mbox{$\text{Range}\widetilde{\mbox{$\rm [Fe/H]$}}(age)$} (bottom left) and one with a positive correlation between the two variables (bottom right). For the bottom left subhalo, even though the RRSF only varies by a small amount ($\sim2$ kpc) across cosmic time, the gradient evolves such that the steepest gradient happens when the new stars are forming in a smaller \mbox{$\rm \text{R}_\text{birth}$} range than when the gradient is weakest. A more extreme example is shown in the bottom right panel, where, in this case, the RRSF varies a lot. Overall, for TNG50 Milky Way and Andromeda-like galaxies whose RRSF varies by $<8$ kpc throughout the disk's lifetime, we find an excellent ability to recover the time evolution of the metallicity gradient from the range in [Fe/H] over age (median correlation of -0.9). When the RRSF varies more throughout the galaxy's lifetime ($>8$ kpc), the median correlation drops down to -0.51. 

\subsection{\edits{How galaxies form affect ability to recover gradient}}\label{sec:gradfeh_correction}

\begin{figure*}
    \centering
     \includegraphics[width=17cm]{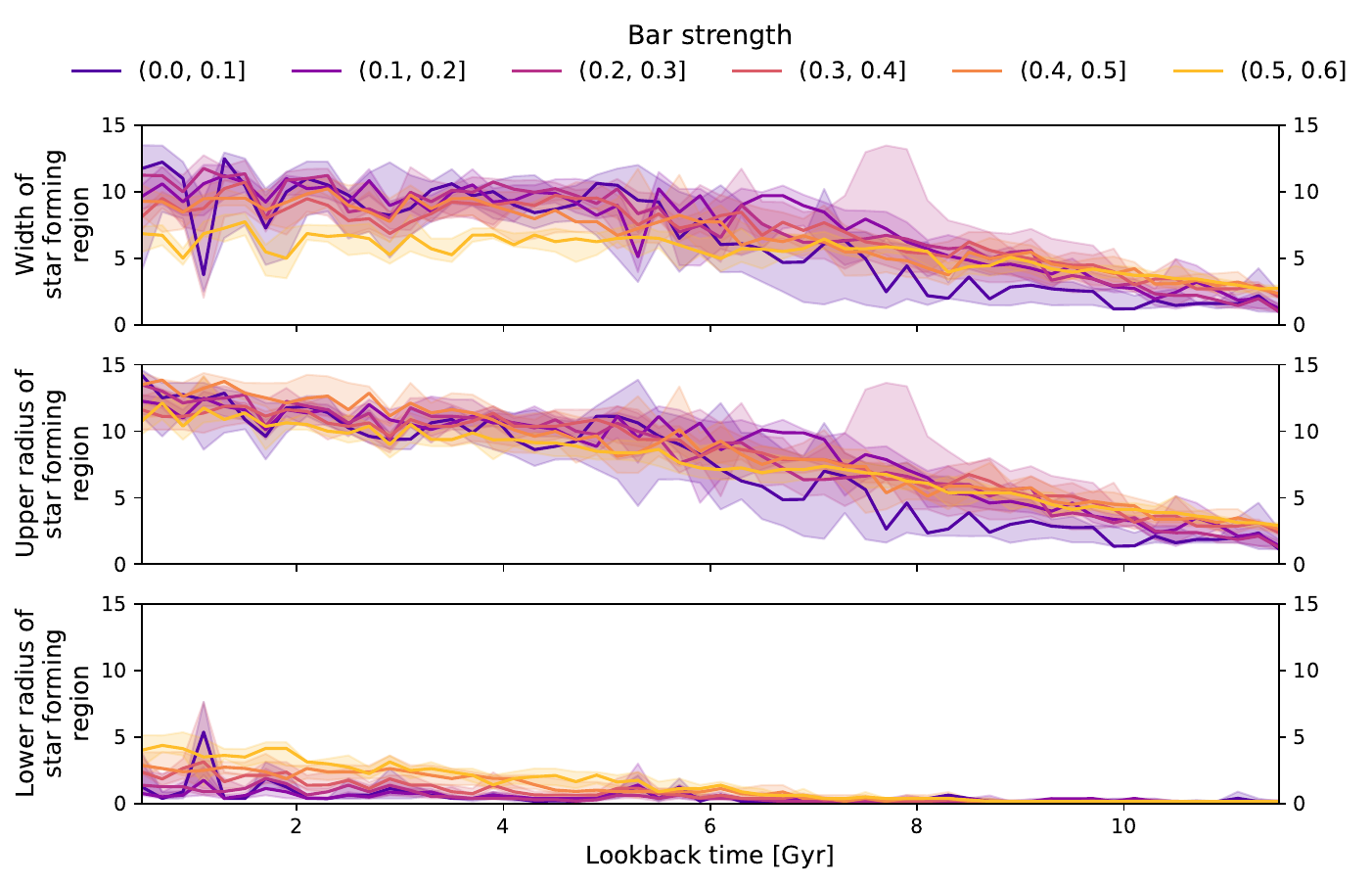}
\caption{\edits{{\bf Top:} Growth of the regions where stars form in the galactic disk over time, separated by bar strength at redshift 0. Weaker-barred galaxies start with stars forming on less than 3 kpc in the center of the disk, and then grow until stars are forming almost throughout the entirety of the disk. Stronger-barred galaxies show only minimal growth in their radial range of star-forming region (RRSF) {\bf Middle:} Upper radial limit of where stars are forming throughout the galaxy at a given moment. Each group of galaxies shows the expected inside-out growth. {\bf Bottom:} 
Lower radial limit of where stars are forming in the galaxy over time. Weaker-barred galaxies tend to form stars consistently in their inner regions, whereas stronger-barred galaxies stop forming stars in the inner few kpc. In each panel, the lines correspond to the median of each group and the shaded region represents the 25-75\% quantiles. To best illustrate the growth of the disk over time, we use age bins of 0.2 Gyr.}}
\label{fig:drTime}
\end{figure*}

\edits{Section \ref{sec:grad_Scatcorr} illustrates that the evolution of the [Fe/H] gradient can be recovered from the range in [Fe/H] across age for stronger-barred galaxies, while weaker-barred galaxies exhibit little to no correlation. The lack of correlation between the [Fe/H] gradient and range is driven by the width of the star forming region varying across time (Fig. \ref{fig:scattGrad_dffR}). The top panel of Figure \ref{fig:drTime} shows how the width of the star forming region changes across time for different barred galaxies. Overall, the weaker-barred galaxies see an increase in the width of the star forming region, where these galaxies begin forming stars on the inner few kpc, and then grow over time until they are forming stars throughout essentially the full disk. This is also illustrated in the middle and bottom panels of the same figure, with the middle panel depicting the growth of the disk while the bottom panel shows that weaker-barred galaxies consistently form stars in the inner regions.}

\edits{Stronger-barred galaxies, on the other hand, show a different evolution. While they appear to have a similar inside-out growth as other galaxies (middle panel Figure \ref{fig:drTime}), stronger-barred galaxies exhibit quenching in the bar region starting about 6-7 Gyr ago. Over time, the bar and lower radius of star formation increase (bottom panel Figure \ref{fig:drTime}), causing the RRSF to remain relatively constant during the disk's evolution, and allowing for a better recovery of the metallicity gradient from the range in [Fe/H] (right panel of Figure \ref{fig:scattGrad_dffR}). This halt in star formation in the inner regions of barred galaxies is seen in other simulations \citep[e.g.][]{Khoperskov2018} and observational data \citep{Geron2024}.}

\section{Results II: recovering the evolution of [Fe/H] at the galactic center} \label{sec:results_feh0}

Now that we have established when the evolution of the metallicity gradient can be recovered from the range in [Fe/H] across age, we move to the other key component in estimating the metallicity profile over time and radius --- the evolution of [Fe/H] in the inner few kpc. In this section, we investigated the ability to recover the evolution of central metallicity (Section \ref{sec:feh0_recover}) and directly tested the assumption that \mbox{$\rm [Fe/H](0, \tau$)}\ is monotonically increasing over time (Section \ref{sec:feh0_mononic}). This section also argues for the use of a projected central metallicity when recovering the galactic disk metallicity profile.

\subsection{Ability to recover the central metallicity with time} \label{sec:feh0_recover}

\begin{figure}
    \centering
     \includegraphics[width=.475\textwidth]{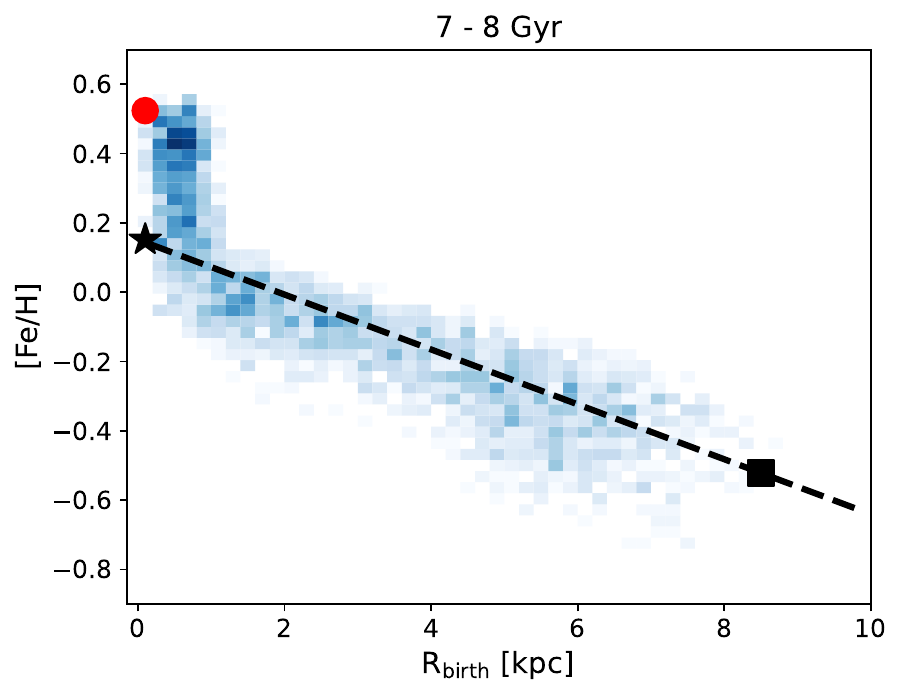}
\caption{Illustrating the recovery of [Fe/H](\mbox{$\rm \text{R}_\text{birth}$} = 0 kpc) for subhalo 553837 for stellar particles 7 - 8 Gyr ago. The projected central metallicity ($\widehat{\text{[Fe/H]}}(\text{R}_\text{birth}$ = 0 kpc); black star) was estimated from the metallicity at a given radius (8.5 kpc; black square) and the metallicity gradient at the current time. The red circle represents the true central metallicity ([Fe/H]($\text{R}_\text{birth} = 0$ kpc), taken as the median [Fe/H]($\mbox{$\rm \text{R}_\text{birth}$} < 1$ kpc). For this scenario, the central metallicity cannot be estimated accurately assuming a linear gradient due to the spike in [Fe/H] in the inner kpc. This galaxy has a weak bar (bar strength of 0.15) and a mass of 3.5 $\times 10^{10}$ M$_\odot$. }
\label{fig:feh0est}
\end{figure}

\begin{figure}
    \centering
     \includegraphics[width=.4\textwidth]{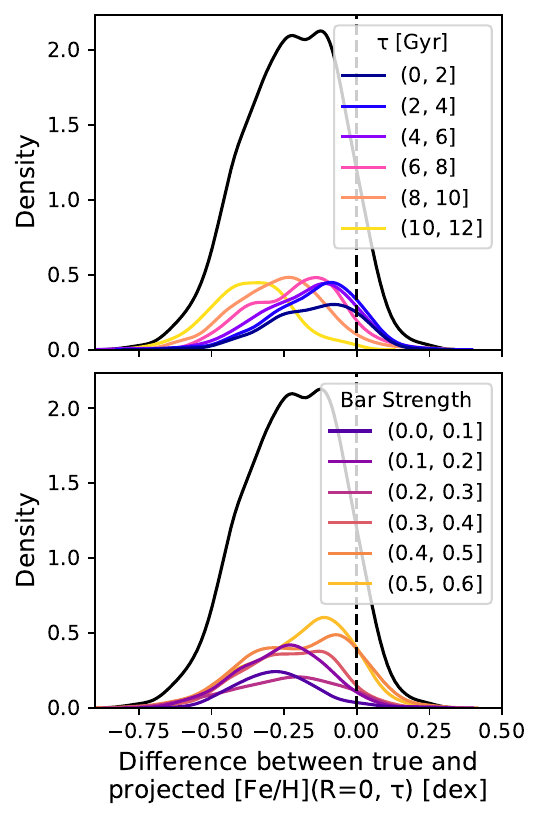}
\caption{Density distribution of the difference between inferred and true [Fe/H] at galactic center separated by \textbf{top}: lookback time $\tau$ and \textbf{bottom}: bar strength. The projected \mbox{$\rm [Fe/H](R = 0$ kpc, $\tau$)}\ is estimated from the [Fe/H] value at a given radius and the current [Fe/H] gradient for a given age of width 1 Gyr (see Figure \ref{fig:feh0est}). As a whole, the value of \mbox{$\rm [Fe/H](R = 0$ kpc, $ \tau$)}\ is underestimated by $\sim 0.2$ dex, with larger differences for weaker barred galaxies and larger lookback times. The bias towards negative values (i.e. \mbox{$\rm [Fe/H](R = 0$ kpc, $ \tau$)}\ is underestimated) is predominantly due to the non-linear spike in [Fe/H] in the inner kpc (see Figure \ref{fig:feh0est} for an example).}
\label{fig:feh0Diff}
\end{figure}

As discussed in Section \ref{sec:method}, estimating birth radii requires knowledge of [Fe/H]($\rm \text{R}_\text{birth}$ = 0 kpc, $\tau$). In recent methods, this quantity is estimated from the upper envelope of the age-metallicity relation for older stars, and extrapolated for younger ($\lesssim 8$ Gyr) stars. Since current samples do not have young stars that were formed in the Galactic center, the boundary condition for this extrapolation is that [Fe/H]($\rm \text{R}_\text{birth}$ = 0 kpc, $\tau = 0$ Gyr) is estimated from the metallicity in the solar neighborhood and the present day gradient of the youngest stars. Figure \ref{fig:corr_feh_rb} illustrated that on average, the galaxies of TNG50 have a fairly linear gradient. However, as these gradients are not completely linear, we wish to test our ability in recovering [Fe/H]($\text{R}_\text{birth}$ = 0 kpc, $\tau$). While the method of \cite{Lu2022_Rb} only uses this technique to recover the boundary condition for the present day, here we investigate it for all age bins that the galaxy has an on average linear gradient to gather a more complete picture. 

Figure \ref{fig:feh0est} illustrates an example of how [Fe/H]($\text{R}_\text{birth}$ = 0 kpc, $\tau$) is estimated for a given period of time. Using the metallicity of a given radius (in this case 8.5 kpc; black square) and the metallicity gradient at that time, we can estimate the projected [Fe/H](\mbox{$\rm \text{R}_\text{birth}$} = 0 kpc) ($\widehat{\text{[Fe/H]}}(\text{R}_\text{birth}$ = 0 kpc); black star). The true central metallicity is taken as the median [Fe/H]($\mbox{$\rm \text{R}_\text{birth}$} < 1$ kpc) of disk stars, and is shown as the red circle. In this example, the estimated central metallicity underpredicts the truth by nearly 0.4 dex due to the large spike in [Fe/H] in the galactic center. Figure \ref{fig:feh0Diff} shows the density distribution of the difference between the true and projected central metallicities across all times and galaxies where the galaxies have stars forming at $\mbox{$\rm \text{R}_\text{birth}$} < 1$ kpc, with the top panel separated by bins of lookback time and the bottom panel separated by bar strength at redshift 0. The differences between the true and projected central metallicity can be quite large (up to 0.5 dex), with a bias towards negative differences. This bias indicates that the projected central metallicity typically underestimates the true central metallicity, which is precisely what was witnessed in Figure \ref{fig:feh0est}. The offset is more noticeable for larger lookback times and weaker barred galaxies, though the underestimation exists for nearly all instances.

The consistency in the projected central metallicity underestimating the true central metallicity illustrates that the spike in the galactic central region's [Fe/H] is a stable feature, at least in TNG50 Milky Way and Andromeda analogues. While the difference can be up to $\sim -0.5$ dex, this underestimation will only affect the recovery of the metallicity evolution for the inner kpc. The remainder of the disk's [Fe/H] profile is fairly well estimated (such as in Figure \ref{fig:feh0est}). In terms of estimating birth radii, the stars in this [Fe/H]-spike region will be given negative birth radii, and thus their birth radii are not able to be recovered with current assumptions. 
Although the centres of the TNG50 galaxies may not be perfectly resolved, keeping in mind the mass resolution of $\rm 8\times10^4\, M_\odot$, the simulations are still likely to capture differences in physical conditions for star formation and its efficiency relative to the rest of the disk~\citep{2021A&A...656A.133Q,2023A&A...673A.147P}. This may result in starbursts, especially at higher redshifts when mergers and interactions are more abundant~\citep{2019A&A...625A..65R}. For instance, the higher pressure in the galactic nuclei is likely to result in a more efficient star formation where the released enriched gas, while being captured in the potential well, recirculated on a shorter time scale, thus resulting in a rapid but local enhancement in the metallicity of newly formed stars. However, such a process can be counteracted by the AGN feedback or other types of star formation quenching; though, it is not the case in many galaxies since we observe continuous star formation in their centres. 



\subsection{Projected [Fe/H] evolution at the galactic center} \label{sec:feh0_mononic}

\begin{figure*}
     \centering
     \includegraphics[width=\textwidth]{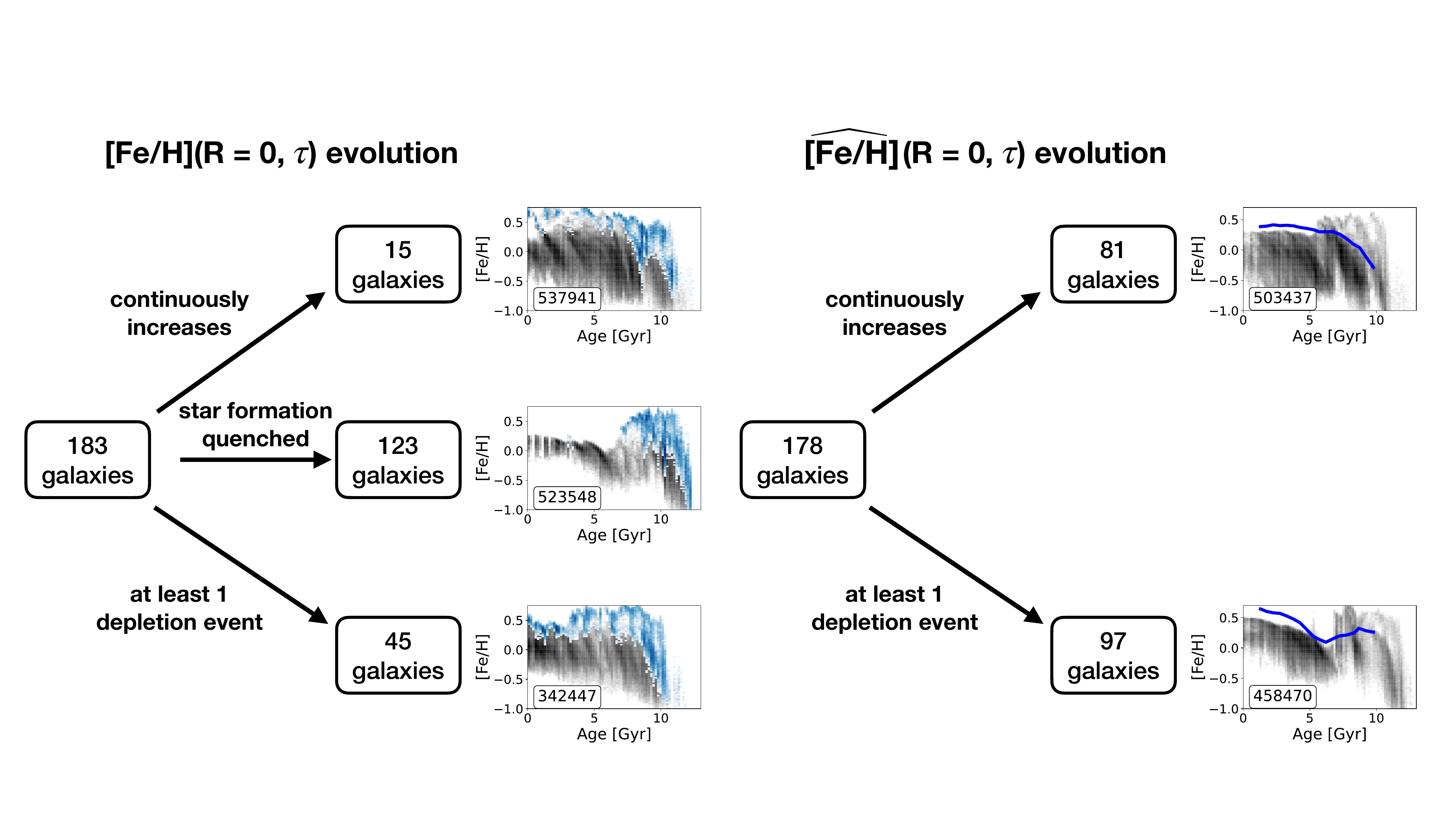}
\caption{Schematic illustrating the distribution of galaxies and how their \textbf{left:} \mbox{$\rm [Fe/H](R = 0$ kpc, $\tau$)}\ and \textbf{right:} $\widehat{\text{[Fe/H]}}(\text{R}_\text{birth}$ = 0 kpc) evolves over time. The evolution of the true central metallicity (left) was estimated by tracing how [Fe/H] in the central kpc of the galactic disk sample enriches over time. Of the 183 galaxies that have enough disk stars born within the inner kpc, only 15 of them are found to have their centers predominantly continuously enrich in [Fe/H] over cosmic time, while 45 of the galaxies are found to have at least one major depletion event in their galactic centers. The remaining 123 galaxies had star formation quench before redshift 0 in their center. An example of each of these scenarios is given, where the density of the age--metallicity relation for our defined disk sample (grey) and stellar particles with  R$_\text{birth} < 1$ kpc (blue) is shown. The evolution of the projected central metallicity (right) was determined by tracing the projected [Fe/H] at $\text{R}_\text{birth}$ = 0 kpc using the metallicity at a given radius and the current metallicity gradient (see Figure \ref{fig:feh0est} for an example). $\widehat{\text{[Fe/H]}}(\text{R}_\text{birth}$ = 0 kpc) traces the central metallicity assuming a linear metallicity gradient, and acts as an anchor point for the disk's gradient, regardless if the true central metallicity spikes in the galactic center. Of the 178 galaxies with enough disk stars to appropriately trace the time evolution of $\widehat{\text{[Fe/H]}}(\text{R}_\text{birth}$ = 0 kpc), we find 81 galaxies show a continuous enrichment at $\text{R}_\text{birth}$ = 0 kpc, while 97 have at least one major depletion event. Examples of these two scenarios are given next to the respective outcome, with the blue line indicating the projected central metallicity, estimated using age bins of 0.5 Gyr. The central metallicity of the Milky Way is assumed to continuously increase in current methods, with potentially only minor deviations from this assumption (see Section \ref{sec:discussion_MWsuccess2}). }
\label{fig:schematic}
\end{figure*}

\begin{figure}
     \centering
     \includegraphics[width=.5\textwidth]{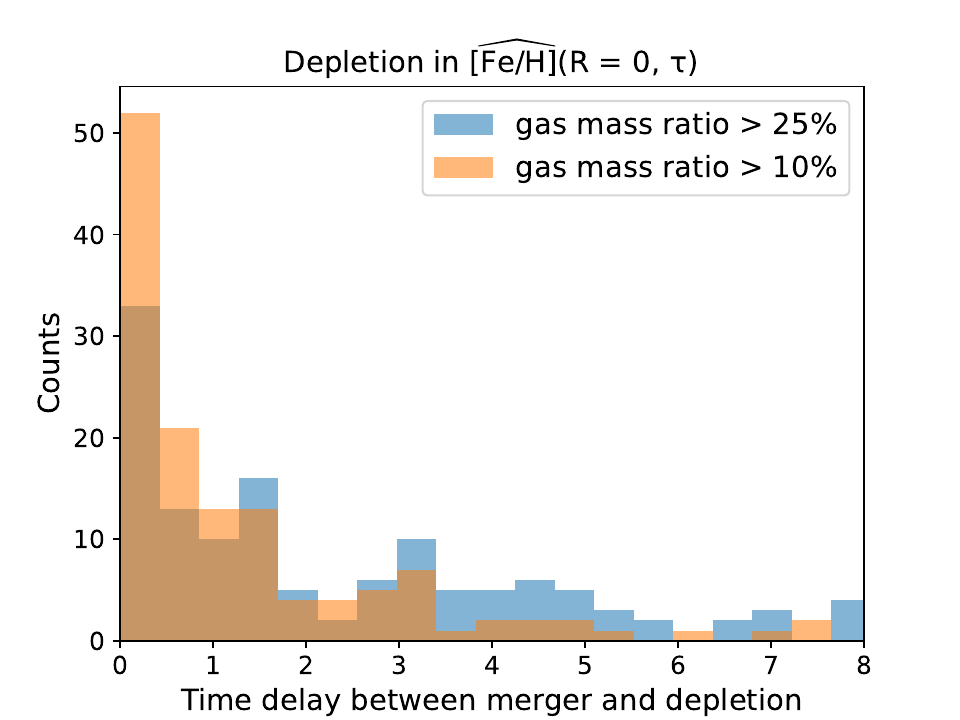}
\caption{Distribution of times between the depletion event in $\widehat{\text{[Fe/H]}}(\text{R}_\text{birth}$ = 0 kpc, $\tau$) and the last major (blue; gas mass ratio $>25\%$) and minor/major (orange; gas mass ratio $>10\%$) mergers. A depletion event is defined as the projected central metallicity decreasing by at least 0.05 dex in 0.5 Gyr, smoothed over 2.5 Gyr. The majority of the time that the projected metallicity in the galactic central region decreases, the galaxy experienced a merger of at least 10\% gas mass fraction within the previous 2 Gyr. This suggests that larger mergers potentially affect the evolution of the metallicity of the galactic center.}
\label{fig:mergerTimes}
\end{figure}

A main assumption in the birth radii method used in \cite{Minchev2018_rbirth, Lu2022_Rb, Ratcliffe2023_enrichment, Ratcliffe2023_chemicalclocks} is that the metallicity at the Galactic center ([Fe/H]($\text{R}_\text{birth}$ = 0 kpc, $\tau$)) is monotonically increasing. However, this assumption can be violated in simulations, where gas-rich mergers cause a dilution in the metallicity of stars forming throughout the host galaxy \citep[e.g.][]{2020_buckchemical, Agertz2021_vintergatanI, Annem2024, Renaud2024}. This assumption also proves false in the majority of the galaxies in our TNG50 Milky Way and Andromeda-like sample, where we find over 90\% of the galaxies in our sample show a non-monotonic evolution in [Fe/H] in the galactic center ($\mbox{$\rm \text{R}_\text{birth}$} < 1$ kpc) to redshift 0. The left schematic of Figure \ref{fig:schematic} illustrates the evolution of the true central metallicity for each galaxy. Starting with 183 galaxies (15 galaxies have too few ages with $\mbox{$\rm \text{R}_\text{birth}$} < 1$ kpc), we find 15 do not show a depletion in \mbox{$\rm [Fe/H](\text{R}_\text{birth} = 0$ kpc, $\tau$)} down to redshift 0, while 45 galaxies show at least one major depletion in [Fe/H] in their galactic centers. The remaining 123 galaxies exhibit quenching at some point in their central kpc, and no stars are formed with  $\text{R}_\text{birth} < 1$ kpc after. The mechanisms of the star formation quenching in the galactic center can be diverse, mainly including morphological~\citep{2009ApJ...707..250M, 2015MNRAS.454.3299R, Khoperskov2018} and AGN-feedback~\citep{2018MNRAS.475..624N, 2018MNRAS.479.4056W,2019MNRAS.490.3234N}, whose relative importance across the sample of our galaxies requires further investigation and is outside the scope of this work. 

Figure \ref{fig:schematic} illustrates that most galaxies have star formation halted in their galactic centers sometime during their evolution. This shows that tracing the upper boundary of [Fe/H] in the age--metallicity plane across all ages will rarely provide the evolution of [Fe/H]($\text{R}_\text{birth}$ = 0 kpc, $\tau$), even with the most complete sample. Additionally, the previous section (Section \ref{sec:feh0_recover}) discussed how the metallicity at the galactic center is not necessarily representative of the metallicity profile in the remainder of the disk, at least for our TNG50 Milky Way and Andromeda-like sample. Thus, we propose not to consider the true central metallicity when deriving the disk's metallicity evolution with time, and hence birth radii, but rather use the projected metallicity $\widehat{\text{[Fe/H]}}(\text{R}_\text{birth}$ = 0 kpc, $\tau$). Focusing on the evolution of the projected metallicity (right schematic in Figure \ref{fig:schematic}), 46\% of the galaxies show a continuously enriching projected central metallicity, and 54\% of the galaxies exhibit at least one major dilution. 

Figure \ref{fig:schematic} informs us that many galaxies experience a dilution in their inner kpc and projected $\widehat{\text{[Fe/H]}}(\text{R}_\text{birth}$ = 0 kpc, $\tau$), and break the assumption of a monotonically enriching galactic center. To investigate potential reasons why, we show the distribution of times between the depletion in $\widehat{\text{[Fe/H]}}(\text{R}_\text{birth}$ = 0 kpc, $\tau$) and when the galaxy experienced its last major (blue) and minor/major (orange) in Figure \ref{fig:mergerTimes}. In the sample of 198 galaxies, we find 131 instances of the central metallicity decreasing by at least 0.05 dex in 0.5 Gyr. This indicates that some galaxies experienced more than one depletion in their central [Fe/H]. Of these drops, 79\% of them had a merger with a gas mass ratio $> 10\%$ in the 2 Gyr prior to the depletion, and 58\% of these instances had a massive merger (gas mass ratio $>25\%$) within the previous 2 Gyr. This shows that there is a correlation between mergers and depletion in $\widehat{\text{[Fe/H]}}(\text{R}_\text{birth}$ = 0 kpc, $\tau$). 

\begin{figure*}
     \sidecaption
     \includegraphics[width=12cm]{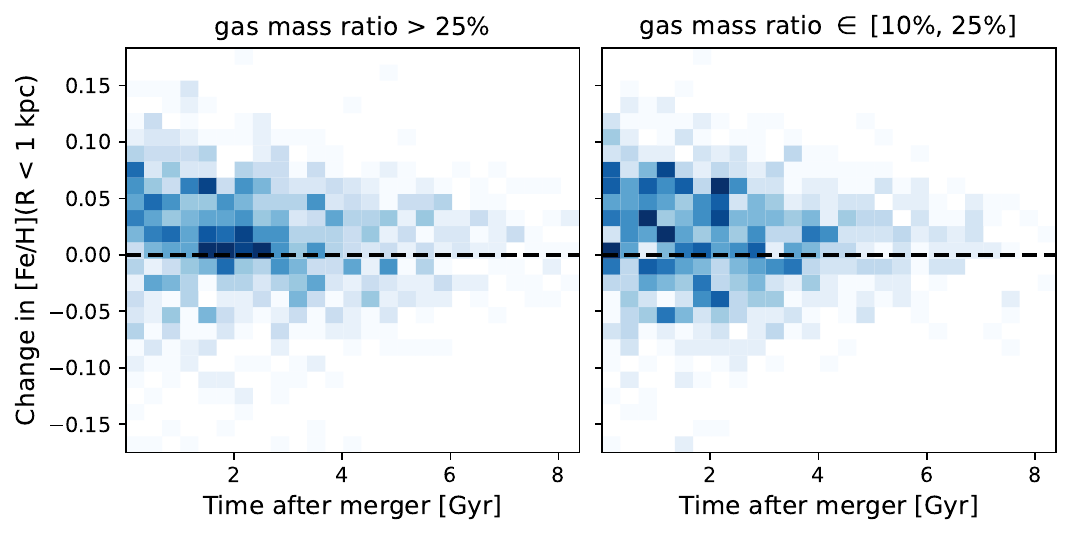}
\caption{Density of the change in $\widehat{\text{[Fe/H]}}(\text{R}_\text{birth}$ = 0 kpc, $\tau$) as a function of time after a (\textbf{left}) massive or (\textbf{right}) minor merger for all galaxies after all massive/minor mergers, until either another massive/minor merger occurs or no more stars form in the central 1 kpc. A positive change in [Fe/H] indicates that the projected metallicity in the central 1 kpc increased, while a negative value indicates that the projected [Fe/H] in the central kpc decreased. The projected metallicity at the galactic center does not necessarily decrease immediately after a merger, and a dilution in $\widehat{\text{[Fe/H]}}(\text{R}_\text{birth}$ = 0 kpc, $\tau$) can take up to a few Gyr before diluting the projected galactic center in [Fe/H]. Even if $\widehat{\text{[Fe/H]}}(\text{R}_\text{birth}$ = 0 kpc, $\tau$) does not dilute, there is a general trend that [Fe/H] enrichment slows down.}
\label{fig:fehchange}
\end{figure*}

Figure \ref{fig:mergerTimes} answers the question of if there was a merger prior to a depletion event. However, this does not necessarily mean that mergers always cause dilutions in $\widehat{\text{[Fe/H]}}(\text{R}_\text{birth}$ = 0 kpc, $\tau$). To answer what happens to the projected central metallicity after a merger, Figure \ref{fig:fehchange} shows the change in $\widehat{\text{[Fe/H]}}(\text{R}_\text{birth}$ = 0 kpc, $\tau$) as a function of time after a merger event for each galaxy after a merger, until either another merger happens or redshift 0. While some galactic centers dilute in projected [Fe/H] immediately after a merger, there is the general trend that it can take up to a few Gyr before $\widehat{\text{[Fe/H]}}(\text{R}_\text{birth}$ = 0 kpc, $\tau$) dilutes. This result suggests that mergers do not necessarily cause the projected central metallicity to dilute, but if there is a dilution, it is not instantaneous. It also shows that mergers can slow down enrichment in the projected central metallicity, even if there is not a dilution. 

\begin{figure*}
     \sidecaption
     \includegraphics[width=12cm]{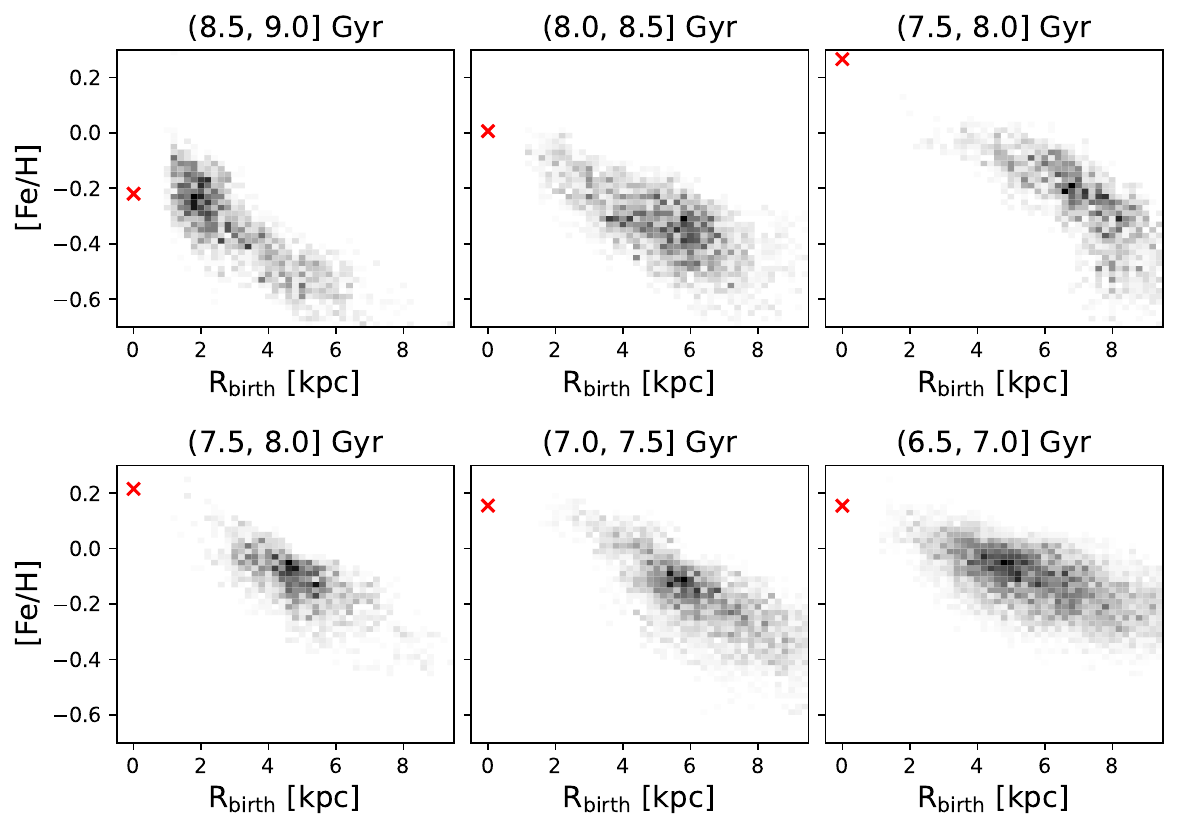}
    \caption{\edits{Two examples of the evolution of $\widehat{\text{[Fe/H]}}(\text{R}_\text{birth}$ = 0 kpc, $\tau$) after a merger with pre-enriched gas. Due to the metal-rich spike in the galactic center (Figure \ref{fig:feh0est}), we remove the inner kpc for visualization purposes. The subhalo in the top row (subhalo 554523) had a major merger at lookback time $\sim9$ Gyr ago which brought in more metal-rich gas and increased the metallicity of the entire disk. This allowed for the projected central metallicity (marked with the red "x") to also increase. The subhalo in the bottom row (subhalo 514829) experienced a metal-rich merger $\sim 8$ Gyr ago. However, instead of increasing the metallicity throughout the entire disk like  in the top row, here only the outer disk had an increase in [Fe/H]. This thus weakened the metallicity gradient, and in turn lowered the projected central metallicity. These examples show that even something as trivial as the metallicity of the merger compared to the main galaxy cannot begin to disentangle when $\widehat{\text{[Fe/H]}}(\text{R}_\text{birth}$ = 0 kpc, $\tau$) increases or decreases after a merger.}}
\label{fig:examples}
\end{figure*}

\edits{In addition to showing that mergers may not cause an instantaneous decrease in the projected central metallicity, Figure \ref{fig:fehchange} also shows that $\widehat{\text{[Fe/H]}}(\text{R}_\text{birth}$ = 0 kpc, $\tau$) can continue to increase after a merger. In our investigations, we find no overarching relationship between merger properties and their effect on the projected metallicity. For instance, when the merger brings in pre-enriched gas, the overall metallicity of the disk can increase, causing an increase in the projected metallicity (top panels of Figure \ref{fig:examples}). However, a merger with pre-enriched gas can also cause the projected central metallicity to decrease (bottom panels of Figure \ref{fig:examples}). This is due to the enriched gas causing the [Fe/H] gradient to weaken, and therefore the extension of the disk to the center is thus lowered. If the merger contains pristine gas, the overall metallicity of the disk can decrease (as seen in e.g. VINTERGATAN; \citealt{Renaud2024}) and/or the gradient of the disk can steepen (as discussed in \citealt{Buck2023}) and therefore create an increase in the projected central metallicity. A detailed paper providing an in depth investigation into what types of mergers cause a depletion or continual increase would provide better insights into the conclusions we draw here.}

\section{Discussion} \label{sec:discussion} 


\subsection{Probability of success in recovering birth radii for disk stars in the Milky Way} \label{sec:discussion_MWsuccess}

\begin{figure*}
     \centering
     \includegraphics[width=\textwidth]{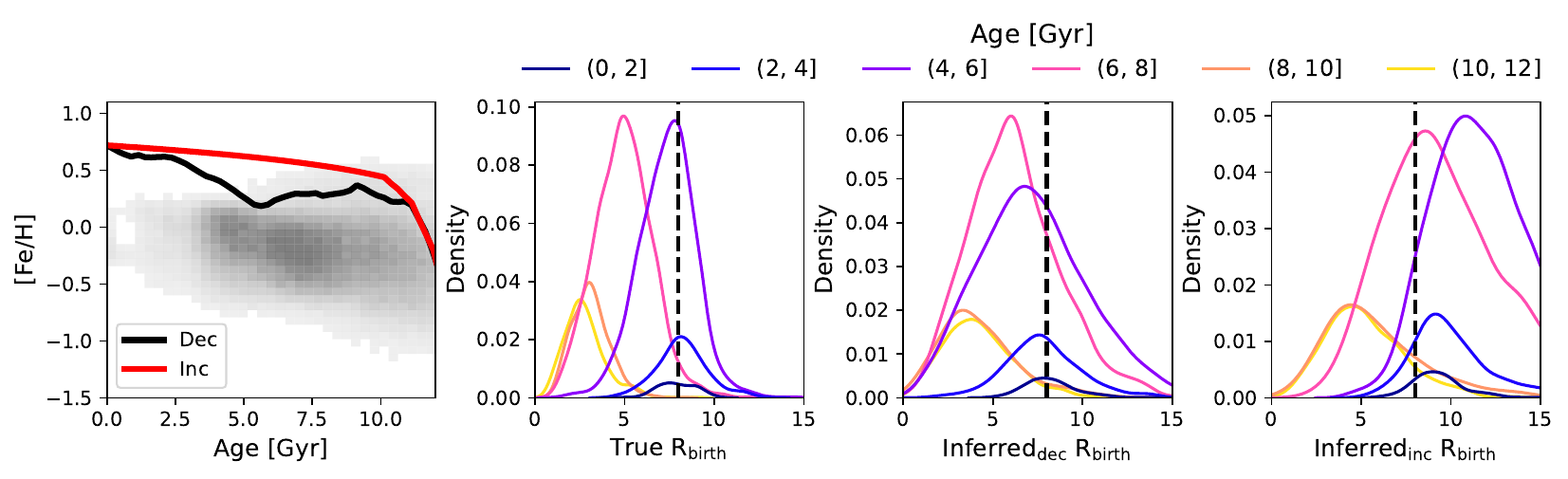}\\

\caption{Ability to recover \mbox{$\rm \text{R}_\text{birth}$} for subhalo 560751 with age and [Fe/H] uncertainties added (15\% and 0.05 dex respectively). \textbf{Left}: The age--metallicity relation for this subhalo with the true evolution of $\widehat{\text{[Fe/H]}}(\text{R}_\text{birth}$ = 0 kpc, $\tau$) (black line) and estimated evolution assuming the projected galactic center is always increasing in [Fe/H] (red line). The \textbf{second}, \textbf{third}, and \textbf{right} panels show the \mbox{$\rm \text{R}_\text{birth}$} distributions for mono-age populations of solar neighborhood ($7.75 \leq R \leq 8.25$) stellar particles for the true \mbox{$\rm \text{R}_\text{birth}$}, \mbox{$\rm \text{R}_\text{birth}$} inferred using the true $\widehat{\text{[Fe/H]}}(\text{R}_\text{birth}$ = 0 kpc, $\tau$) ($\text{Inferred}_\text{dec}$ \mbox{$\rm \text{R}_\text{birth}$}), and \mbox{$\rm \text{R}_\text{birth}$} inferred using the monotonically increasing $\widehat{\text{[Fe/H]}}(\text{R}_\text{birth}$ = 0 kpc, $\tau$) ($\text{Inferred}_\text{inc}$ \mbox{$\rm \text{R}_\text{birth}$}) respectively. Inappropriately assuming a monotonically increasing metallicity abundance in the galactic center provides \mbox{$\rm \text{R}_\text{birth}$} distributions that are non-physical and do not match the true distributions, whereas using the true $\widehat{\text{[Fe/H]}}(\text{R}_\text{birth}$ = 0 kpc, $\tau$) evolution provides a more accurate trend in how \mbox{$\rm \text{R}_\text{birth}$} decreases with increasing age. Examining the $\rm \text{R}_\text{birth}$ distributions of mono-age populations in the solar neighborhood is thus a necessity in constraining  $\widehat{\text{[Fe/H]}}(\text{R}_\text{birth}$ = 0 kpc, $\tau$) evolution.} 
\label{fig:Rb_est}
\end{figure*}

While hydrodynamical cosmological simulations and Galactic chemical evolution models have made great advances in understanding the Milky Way's formation and evolution, there are still discrepancies that only exploring the Milky Way itself will answer. Knowledge of stellar birth radii in the Milky Way disk provides a useful avenue in inferring the disk's evolution, from understanding the strength of migration \citep[e.g.][]{Frankel2018} and mergers on the Galactic disk \citep{Lu2022_Rb}, inferring the chemical evolution of the disk \citep{Ratcliffe2023_enrichment, Ratcliffe2023_chemicalclocks, Wang2023}, and constraining chemo-kinematic relations \citep{Minchev2018_rbirth}. However, we cannot directly measure \mbox{$\rm \text{R}_\text{birth}$}, and need to infer it using some method \citep{Minchev2018_rbirth, Frankel2018, Lu2022_Rb, Wang2023}.

The focus of this paper is to investigate the recent method from \cite{Lu2022_Rb} (whose assumptions are quite similar to previous methods and necessary for \mbox{$\rm \text{R}_\text{birth}$} recovery) with TNG50 Milky Way and Andromeda-like galaxies, and understand its success in the context of the Milky Way itself. The method uses Equation \ref{eqn:rb}, where \mbox{$\rm \nabla [Fe/H](\tau)$} is estimated from the range in [Fe/H] over different age bins (Section \ref{sec:grad_Scatcorr}) and \mbox{$\rm [Fe/H](R = 0$ kpc, $\tau$)}\ is estimated from the upper envelope of older stars (when the sample is expected to contain stars that formed in the inner few kpc of the Galaxy) and modeled as continuously enriching in [Fe/H] until the present day, with the boundary condition estimated using the current metallicity and gradient of the youngest stars in the solar neighborhood (Section \ref{sec:feh0_recover}). Overall, we find that the individual aspects of this method seem like they can be reasonably extended to the Milky Way disk. Most galaxies in the sample show that the metallicity gradient can be recovered from the range in [Fe/H] across age (Figure \ref{fig:scattGrad_hist}). The galaxies that fail this condition tend to have weaker bars compared to the stronger one of the Milky Way (at least in the context of TNG50; \citealt{Khoperskov2023}) and have large variations in the range of their star forming region over time ($>$ 8 kpc). While the median correlation between \rm \mbox{$\text{Range}\widetilde{\mbox{$\rm [Fe/H]$}}(age)$} and \mbox{$\rm \nabla [Fe/H](\tau)$} is -0.85 for galaxies whose star-forming region can vary in size by up to 8 kpc, a more accurate method would account for the growth of the Galactic disk \citep[e.g.][]{Molla2019}. 

While the metallicity gradient provides the relative [Fe/H] across the disk, \mbox{$\rm [Fe/H](R = 0$ kpc, $ \tau$)}\ acts as an anchor in determining the precise [Fe/H] as a function of radius, even if there is no star formation in the galactic center over the previous few Gyr. Section \ref{sec:results_feh0} discussed the importance of using the projected central metallicity ($\widehat{\text{[Fe/H]}}(\text{R}_\text{birth}$ = 0 kpc, $\tau$)) due to the possibly non-representative [Fe/H] profile in the inner few kpc. Section \ref{sec:feh0_mononic} suggests that the assumption that $\widehat{\text{[Fe/H]}}(\text{R}_\text{birth}$ = 0 kpc, $\tau$) is continuously enriching in metallicity is potentially violated for the average Milky Way and Andromeda-like galaxy, since we find just over 50\% of the galaxies in our sample have at least one dilution in the projected [Fe/H] in their galactic centers at some point during their evolution after the disk formed. However, the vast majority of the times when the galactic centers see a depletion in projected [Fe/H] are within 2 Gyr of a merger with gas mass ratio $>10\%$ (Figure \ref{fig:mergerTimes}). This suggests that the Milky Way's Galactic center probably did not see a decrease in [Fe/H] with time due to Sagittarius. Though, it does raise the question of the validity of $\widehat{\text{[Fe/H]}}(\text{R}_\text{birth}$ = 0 kpc, $\tau$) monotonically increasing after the GSE merger event. 

\subsubsection{Incorrect projected central metallicity assumption gives non-physical birth radii}\label{sec:discussion_MWsuccess2}

An additional constraint in the \mbox{$\rm \text{R}_\text{birth}$} methods from \cite{Minchev2018_rbirth, Lu2022_Rb} is that the birth radii distributions for mono-age populations in the solar neighborhood need to be physically meaningful (see Figure A3 in \citealt{Lu2022_Rb}). In particular, the youngest stars in the solar neighborhood are expected to be locally born, while the oldest stars are expected to be born more inwards \citep{Minchev2013, Netopil2022, Agertz2021_vintergatanI, Carrillo2023}. Therefore, a natural test to determine the effect of incorrectly assuming a monotonically increasing $\widehat{\text{[Fe/H]}}(\text{R}_\text{birth}$ = 0 kpc, $\tau$) is to see how the \mbox{$\rm \text{R}_\text{birth}$} distributions of mono-age populations of stellar particles near 8 kpc vary with this assumption. Figure \ref{fig:Rb_est} shows this test for one of our TNG50 Milky Way and Andromeda-like galaxies (subhlao 560751). This galaxy is like the Milky Way in many ways: it has a stronger bar strength (0.61), M$_*(<30$kpc) = $4.12 \times 10^{10}$ M$_\odot$, and it appears to have no stars forming in the inner few kpc for the last $\sim5$ Gyr. The projected metallicity in the galactic center of this galaxy does not monotonically increase, as illustrated by the black line in the age--metallicity relation shown in the left panel of Figure \ref{fig:Rb_est}. The red line in the same panel shows what central metallicity evolution is assumed for the Milky Way in \cite{Lu2022_Rb, Ratcliffe2023_enrichment, Ratcliffe2023_chemicalclocks}; the upper [Fe/H] envelope for older ages and a log function for younger ages with the boundary condition that the $\widehat{\text{[Fe/H]}}(\text{R}_\text{birth}$ = 0 kpc, $\tau$) is estimated from the present day gradient of the youngest stars and the metallicity at a given radius. We estimated \mbox{$\rm \text{R}_\text{birth}$} in two different ways for this galaxy --- using the true, declining $\widehat{\text{[Fe/H]}}(\text{R}_\text{birth}$ = 0 kpc, $\tau$) (black line from left panel of Figure \ref{fig:Rb_est}) which we call Inferred$_\text{dec}$ \mbox{$\rm \text{R}_\text{birth}$} and using the estimated, inclining $\widehat{\text{[Fe/H]}}(\text{R}_\text{birth}$ = 0 kpc, $\tau$) (red line from same figure) which we call Inferred$_\text{inc}$ \mbox{$\rm \text{R}_\text{birth}$}. For both ways, we used the gradient estimated from \rm \mbox{$\text{Range}\widetilde{\mbox{$\rm [Fe/H]$}}(age)$} in Equation \ref{eqn:rb}. To make the best connections between the simulation and the Milky Way, we estimated \mbox{$\rm \text{R}_\text{birth}$} after redrawing new age and [Fe/H] parameters for each stellar particle by using uncertainties of 15\% and 0.05 dex respectively. These errors are on par with the measurement uncertainties of large spectroscopic surveys \citep{Jonsson2020, Xiang2019_lamost, Queiroz2023_SH, Anders2023_ages}.

The other panels of Figure \ref{fig:Rb_est} illustrate the \mbox{$\rm \text{R}_\text{birth}$} distributions of mono-age populations for stellar particles within $7.75 \leq R \leq 8.25$ kpc for the true \mbox{$\rm \text{R}_\text{birth}$} (second panel), Inferred$_\text{dec}$ \mbox{$\rm \text{R}_\text{birth}$} (third panel), and Inferred$_\text{inc}$ \mbox{$\rm \text{R}_\text{birth}$} (right panel). The distributions of the true \mbox{$\rm \text{R}_\text{birth}$} show clear signs of radial migration and inside-out formation with the oldest stars forming $\mbox{$\rm \text{R}_\text{birth}$}<5$ kpc and the younger stars forming in the 8 kpc region. Recovering \mbox{$\rm \text{R}_\text{birth}$} using the estimated gradient and declining $\widehat{\text{[Fe/H]}}(\text{R}_\text{birth}$ = 0 kpc, $\tau$) (Inferred$_\text{dec}$ \mbox{$\rm \text{R}_\text{birth}$}) shows similar trends with age. Assuming a monotonically increasing $\widehat{\text{[Fe/H]}}(\text{R}_\text{birth}$ = 0 kpc, $\tau$) provides incorrect distributions, where now the stars forming $4-8$ Gyr ago are estimated to have formed beyond 8 kpc, with a large percentage supposedly forming beyond 15 kpc. Additionally, the youngest ($\leq 2$ Gyr) are found to be born at $\sim 10$ kpc, implying these stars migrated inwards 2 kpc in less than 2 Gyr. Similarly wrong results still exist even when we change the strength of the gradient (see Figure \ref{fig:changeMaxGrad} in the appendix), which has been shown to affect the \mbox{$\rm \text{R}_\text{birth}$} distributions of mono-age populations \citep{Lu2022_Rb}. This finding holds good news for extending the \mbox{$\rm \text{R}_\text{birth}$} method to the Milky Way disk. The \mbox{$\rm \text{R}_\text{birth}$} distributions of mono-age populations for solar neighborhood stars in the Milky Way have been found to show a clear decrease in \mbox{$\rm \text{R}_\text{birth}$} with increasing age (see Fig. 6 in \citealt{Lu2022_Rb} and Fig. 3 in \citealt{Ratcliffe2023_enrichment}), much like the distributions found in the TNG50 galaxy for the \mbox{$\rm \text{R}_\text{birth}$} estimated using the decreasing $\widehat{\text{[Fe/H]}}(\text{R}_\text{birth}$ = 0 kpc, $\tau$); the distributions for Milky Way disk stars does not resemble the distributions found when incorrectly assuming the galactic center's metallicity monotonically increases. This finding strongly suggests that the metallicity evolution of the Milky Way's inner few kpc continuously enriches in [Fe/H], with potentially only minor deviations. 

\subsubsection{Assuming Milky Way-like scatter about the metallicity gradient}\label{sec:smallScatt}

\begin{figure*}
     \sidecaption
     \includegraphics[width=12cm]{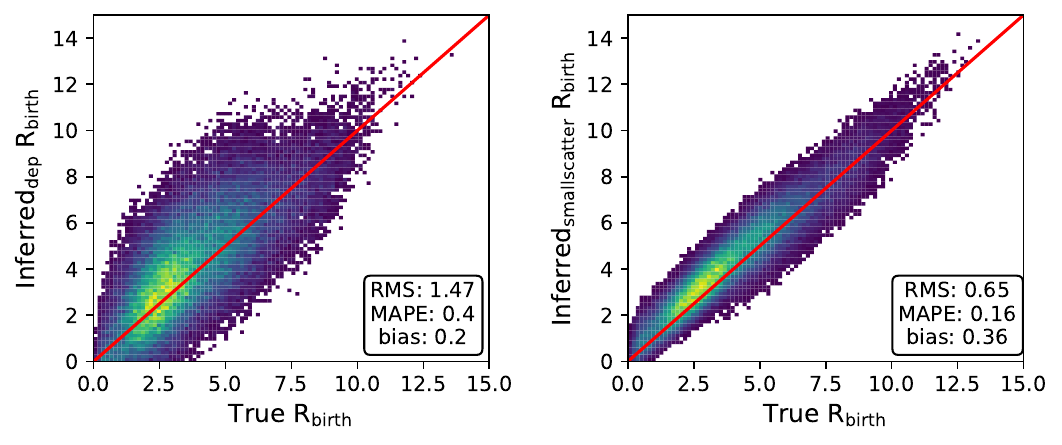}
\caption{Inferred \mbox{$\rm \text{R}_\text{birth}$} versus true \mbox{$\rm \text{R}_\text{birth}$} for subhalo 560751 using the method from \cite{Lu2022_Rb} with age and [Fe/H] uncertainties added of 15\% and 0.05 dex respectively. \mbox{$\rm \text{R}_\text{birth}$} are estimated using the true evolution of \mbox{$\rm [Fe/H](R = 0$ kpc, $ \tau$)}\ (black line in Fig \ref{fig:Rb_est}) and \mbox{$\rm \nabla [Fe/H](\tau)$} recovered from \rm \mbox{$\text{Range}\widetilde{\mbox{$\rm [Fe/H]$}}(age)$}. The difference between the left and right panels is that stellar [Fe/H] values in the TNG50 galaxy were adapted to better match the dispersion about the metallicity gradient as seen in the Milky Way's H II regions in the right panel. This illustrates the high probability of success in estimating \mbox{$\rm \text{R}_\text{birth}$} in the Milky Way disk.} 
\label{fig:Rb_estCompare}
\end{figure*}

The third panel of Figure \ref{fig:Rb_est} shows that even though we are able to correctly recover the trend between age and \mbox{$\rm \text{R}_\text{birth}$} for solar neighborhood stars, the recovered distributions are wider than the truth. This is in part due to the scatter seen in the simulation about the linear metallicity gradient, which we find to be 0.14 dex across all ages for subhalo 560751. This scatter can be explained by strong radial migration, which has been shown to be unrealistically strong in TNG50 simulated galaxies \citep{WangLilly2023}, and may be due to the rapid migration from feedback in simulations \citep{ElBadry2016}. To account for this effect, we now adjust the dispersion about the linear birth metallicity gradient for every age bin (1 Gyr) in TNG50 galaxy 560751 to be on the same order as that seen in H II regions of the Milky Way (0.07 dex; \citealt{Esteban2017, ArellanoCordova2021}). Figure \ref{fig:Rb_estCompare} shows how much more successful the \mbox{$\rm \text{R}_\text{birth}$} method is when the dispersion is taken to be more Milky Way-like. The left panel shows the inferred vs true birth radii when \mbox{$\rm \text{R}_\text{birth}$} is estimated using Equation \ref{eqn:rb} with the estimated \mbox{$\rm \nabla [Fe/H](\tau)$} and the decreasing $\widehat{\text{[Fe/H]}}(\text{R}_\text{birth}$ = 0 kpc, $\tau$) (black line in Figure \ref{fig:Rb_est}). Even in the regime of extra migration from feedback, we are able to recover \mbox{$\rm \text{R}_\text{birth}$} to within 1.34 kpc. When the [Fe/H] dispersion is adjusted to follow that of the Milky Way, the ability to recover \mbox{$\rm \text{R}_\text{birth}$} increases to within 0.67 kpc. This value is on-par with results using higher resolution simulations in recovering \mbox{$\rm \text{R}_\text{birth}$} (\citealt{Lu2022_Rb}). The positive bias implies that the method overestimates the birth radius of a star on average by 0.42 kpc, which we find shrinks with increasing precision in age and [Fe/H].  

\subsection{Do wiggles happen at same age}

When calculating stellar birth radii for APOGEE DR17 \citep{apogeeDR17, Majewski2017} red giants, \cite{Ratcliffe2023_enrichment} found that a given birth radius did not necessarily continuously enrich in [Fe/H] over time (see the right panel of their Fig. 2). These ``wiggles" were seen to happen at the same time for each birth radius (in their case, $\sim4$ and 6 Gyr) with the outer birth radii having the larger fluctuation in [Fe/H]. This structure is caused when \mbox{$\rm [Fe/H](R = 0$ kpc, $ \tau$)}\ minimally varies when the metallicity gradient fluctuates (i.e. flattens, steepens, and then flattens again). While Section \ref{sec:feh0_mononic} showed that most galaxies in our sample do not monotonically increase in [Fe/H] at their center, we still wish to understand if these ``wiggles" do in fact happen at the same time for each birth location, or if the outer (or inner) radii feel this effect first.

\begin{figure*}
     \sidecaption
     \includegraphics[width=12cm]{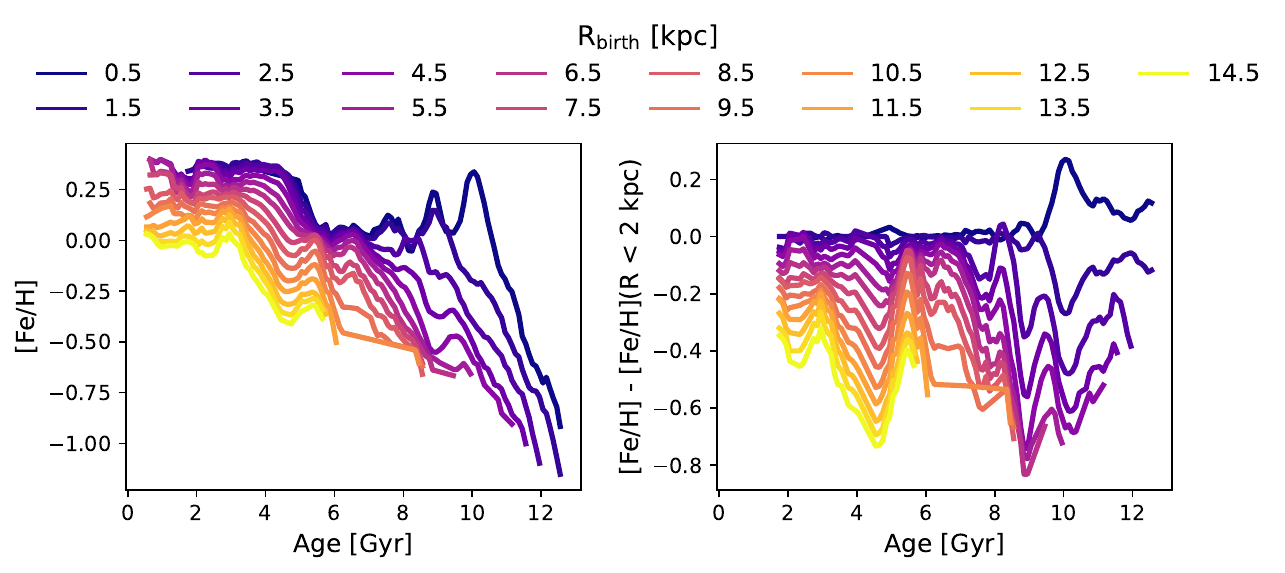}
\caption{\textbf{Left column}: Evolution of mono-\mbox{$\rm \text{R}_\text{birth}$} populations in the age--metallicity relation for a TNG50 Milky Way and Andromeda-like galaxy (subhalo 402555) whose mono-\mbox{$\rm \text{R}_\text{birth}$} populations exhibit ``wiggling" behavior, where each mono-\mbox{$\rm \text{R}_\text{birth}$} population did not experience a monotonic enrichment in [Fe/H]. \textbf{Right column}: [Fe/H] abundance of mono-\mbox{$\rm \text{R}_\text{birth}$} populations relative to [Fe/H](R $<2$ kpc) across age. Looking at the [Fe/H] abundance relative to the inner few kpc at a given age allows the structure of the wiggles to become more clear. Most of the galaxies that exhibit ``wiggling" structure in this plane follow similar trends to this galaxy --- each \mbox{$\rm \text{R}_\text{birth}$} sees a depletion in [Fe/H] at the same time as also seen using  \mbox{$\rm \text{R}_\text{birth}$} methods \citep[see Fig. 2 in][]{Ratcliffe2023_enrichment}.}
\label{fig:wiggles}
\end{figure*}

The left column of Figure \ref{fig:wiggles} shows the age--metallicity relation for mono-\mbox{$\rm \text{R}_\text{birth}$} populations for a galaxy from our sample whose mono-\mbox{$\rm \text{R}_\text{birth}$} populations do not continuously enrich in [Fe/H] and ``wiggle". Due to the variability in \mbox{$\rm [Fe/H](R = 0$ kpc, $ \tau$)}, examining the extent of these wiggles is nontrivial. Therefore, to better match the conditions assumed for the Milky Way disk during the time of gradient fluctuations in \cite{Ratcliffe2023_enrichment}, we also show the metallicity of a mono-\mbox{$\rm \text{R}_\text{birth}$} population relative to the metallicity of the inner 2 kpc in the right column of Figure \ref{fig:wiggles}. Once [Fe/H](R, $\tau$) is put relative to the metallicity of the inner few kpc, the wiggles become much more apparent. It is easier to see that the wiggles happen at the same time, similar to what what shown in \cite{Ratcliffe2023_enrichment}. Most other galaxies that exhibit wiggles follow similar trends. The wiggle at 4 Gyr may be due to two metal-poor mergers that merged $\sim5-6$ Gyr ago (gas merger ratios of 4\% and 10\%), while the dramatic increase in [Fe/H] in the outer disk at $\sim 3$ Gyr may be due to the few small metal-rich mergers at $\sim 4-5$ Gyr. Similar trends can also be seen in the NIHAO-UHD simulation suite \citep{Buck2023}.

\section{Conclusions} \label{sec:conclusion}

This work explores in detail the validity of assumptions needed to estimate [Fe/H](R, $\tau$) and \mbox{$\rm \text{R}_\text{birth}$} using the sample of TNG50 Milky Way and Andromeda analogues, and discusses the implications of these assumptions in the context of the Milky Way. Knowledge of the evolution of the metallicity profile with radius, in addition to stellar \mbox{$\rm \text{R}_\text{birth}$}, opens up a new and powerful avenue to explore the Milky Way's formation path, and can provide necessary constraints on chemo-dynamical evolution. Our key findings are: 

\begin{itemize}
    \item Most Milky Way and Andromeda-like  galaxies in TNG50 have a good correlation between [Fe/H] and \mbox{$\rm \text{R}_\text{birth}$} across time (Figure \ref{fig:corr_feh_rb}). We find that most of the galaxies that do not exhibit a tight correlation either have too shallow of a gradient (which causes a low correlation) or seemingly had their pre-existing disk destroyed by late massive mergers.
    
    \item We can recover the time evolution of the metallicity gradient for most TNG50 Milky Way and Andromeda-like galaxies in our sample from the range in [Fe/H] across age, with a dependency on the bar strength at redshift zero (Figure \ref{fig:scattGrad_hist}). The galaxies that show no correlation between \mbox{$\rm \nabla [Fe/H](\tau)$} and \rm \mbox{$\text{Range}\widetilde{\mbox{$\rm [Fe/H]$}}(age)$} have a large variability in the radial range of their star-forming region over time. The galaxies in our sample with weaker bars (bar strength $<0.2$) predominantly have larger variations in their star-forming region, while the stronger barred galaxies have less variation on average.
    
    \item \mbox{$\rm [Fe/H](R = 0$ kpc, $ \tau$)}\ is not necessarily representative of the metallicity profile of the remainder of the disk due to, for example, spikes in [Fe/H] in the inner few kpc (Figures \ref{fig:feh0est} and \ref{fig:feh0Diff}). The differences are smaller for smaller lookback times and galaxies with stronger bars. We propose instead that recovering the evolution of the metallicity profile of the galactic disk should use a projected central metallicity ($\widehat{\text{[Fe/H]}}(\text{R}_\text{birth}$ = 0 kpc, $\tau$)) that captures the trend of the disk and ignores non-linearities in the center.
    
    \item The evolution of $\widehat{\text{[Fe/H]}}(\text{R}_\text{birth}$ = 0 kpc, $\tau$) is non-monotonic for $\sim55\%$ of the galaxies in our sample (Figure \ref{fig:schematic}). 79\% of these dilutions in [Fe/H] in the galactic center happened within 2 Gyr of a merger with gas mass ratio $>10\%$ (Figure \ref{fig:mergerTimes}).
    
    \item Requiring \mbox{$\rm \text{R}_\text{birth}$} distributions of mono-age populations to remain physically meaningful constrains the assumption of a monotonically increasing $\widehat{\text{[Fe/H]}}(\text{R}_\text{birth}$ = 0 kpc, $\tau$) (Figure \ref{fig:Rb_est}). If the projected metallicity at the galactic center did not continuously increase, then assuming so would imply non-physical \mbox{$\rm \text{R}_\text{birth}$} distributions, with too many stars needing to migrate inwards many kpc in a short time.
\end{itemize}

Our results show that overall, the assumptions required to estimate [Fe/H](R, $\tau$) and stellar \mbox{$\rm \text{R}_\text{birth}$} should hold reasonably well for Milky Way and Andromeda-like disks. The most uncertain assumption is the monotonically enriching $\widehat{\text{[Fe/H]}}(\text{R}_\text{birth}$ = 0 kpc, $\tau$), where we find about half of the galaxies in our sample did not follow this trend. However, our work in Section \ref{sec:discussion_MWsuccess2} suggests that this assumption should hold in the Milky Way, as the \mbox{$\rm \text{R}_\text{birth}$} distributions of mono-age populations in the solar neighborhood are meaningful for the Milky Way disk. Looking in detail at a Milky Way-like galaxy (strong bar, similar mass, little to no star formation in the inner disk for the most recent ages), we expect to be able to recover \mbox{$\rm \text{R}_\text{birth}$} to within 1 kpc for the Milky Way disk. As age and [Fe/H] measurement uncertainties continue to decrease with future surveys, the ability to estimate \mbox{$\rm \text{R}_\text{birth}$} will improve. The increasing precision will also reveal any structure that is currently masked due to observational errors \citep{Renaud2021_vintergatanII}, which should help constrain the recovery of the metallicity profile and stellar birth radii.

\edits{While there still exists uncertainty in chemical yields and simulation prescription \citep{2021BuckHD_chemEnrich}, the [Fe/H] gradients in TNG50 are able to reproduce observed MANGA gradients (\citealt{Lian2023}). Despite having different subgrid physics and implementation, other simulations such as VINTERGATAN (\citealt{Agertz2021_vintergatanI, Renaud2021_vintergatanII, Renaud2021_vintergatanIII}) and NIHAO-UHD (\citealt{2020_buckchemical,2020buck_NIHAO-UHD}) are also able to reproduce well global chemical abundance trends such as the chemical bimodality. Therefore, we believe our findings are applicable to observational data and not strongly affected by specific chemical enrichment (not including the dependence on bar strength discovered in this work). Other works have shown that the metallicity gradient can be recovered from the range in [Fe/H] for a few Milky Way-like galaxies using 2 different simulation suites (NIHAO-UHD and HESTIA; \citealt{Lu2022_Rb}) and additionally on Large Magellanic Cloud-like galaxies using NIHAO-UHD (\citealt{Lu2024_RbLMC}), thus showing the metallicity gradient-range correlation is not an artifact of TNG50's prescription. As discussed in Section \ref{sec:feh0_recover}, the metallicity profile shows a spike in [Fe/H] near the galactic center. However, as discussed, this drastic increase may be a result of TNG-specific evolution as it does not seem to appear consistently across other simulation suites (e.g. \citealt{Renaud2024} and Figure 3 in \citealt{Lu2024_RbLMC}). The main conclusion of Section \ref{sec:feh0_recover} still holds true though, that regardless of what the metallicity profile looks like at the true galactic center, a projected central metallicity is best to estimate stellar birth radii of disk stars.  }

\section*{Acknowledgements}
 
The authors wish to thank Annalisa Pillepich for her helpful discussions and the Illustris TNG collaboration for making their simulations publicly available, along with an active discussion forum. B.R. and I.M. acknowledge support by the Deutsche Forschungsgemeinschaft under the grant MI 2009/2-1. 

\bibliography{Ratcliffe}{}

\begin{thebibliography}{102}
\expandafter\ifx\csname natexlab\endcsname\relax\def\natexlab#1{#1}\fi

\bibitem[{{Abdurro'uf} {et~al.}(2022){Abdurro'uf}, {Accetta}, {Aerts}, {Silva Aguirre}, {Ahumada}, {Ajgaonkar}, {Filiz Ak}, {Alam}, {Allende Prieto}, {Almeida}, {Anders}, {Anderson}, {Andrews}, {Anguiano}, {Aquino-Ort{\'\i}z}, {Arag{\'o}n-Salamanca}, {Argudo-Fern{\'a}ndez}, {Ata}, {Aubert}, {Avila-Reese}, {Badenes}, {Barb{\'a}}, {Barger}, {Barrera-Ballesteros}, {Beaton}, {Beers}, {Belfiore}, {Bender}, {Bernardi}, {Bershady}, {Beutler}, {Bidin}, {Bird}, {Bizyaev}, {Blanc}, {Blanton}, {Boardman}, {Bolton}, {Boquien}, {Borissova}, {Bovy}, {Brandt}, {Brown}, {Brownstein}, {Brusa}, {Buchner}, {Bundy}, {Burchett}, {Bureau}, {Burgasser}, {Cabang}, {Campbell}, {Cappellari}, {Carlberg}, {Wanderley}, {Carrera}, {Cash}, {Chen}, {Chen}, {Cherinka}, {Chiappini}, {Choi}, {Chojnowski}, {Chung}, {Clerc}, {Cohen}, {Comerford}, {Comparat}, {da Costa}, {Covey}, {Crane}, {Cruz-Gonzalez}, {Culhane}, {Cunha}, {Dai}, {Damke}, {Darling}, {Davidson}, {Davies}, {Dawson}, {De Lee}, {Diamond-Stanic}, {Cano-D{\'\i}az}, {S{\'a}nchez},
  {Donor}, {Duckworth}, {Dwelly}, {Eisenstein}, {Elsworth}, {Emsellem}, {Eracleous}, {Escoffier}, {Fan}, {Farr}, {Feng}, {Fern{\'a}ndez-Trincado}, {Feuillet}, {Filipp}, {Fillingham}, {Frinchaboy}, {Fromenteau}, {Galbany}, {Garc{\'\i}a}, {Garc{\'\i}a-Hern{\'a}ndez}, {Ge}, {Geisler}, {Gelfand}, {G{\'e}ron}, {Gibson}, {Goddy}, {Godoy-Rivera}, {Grabowski}, {Green}, {Greener}, {Grier}, {Griffith}, {Guo}, {Guy}, {Hadjara}, {Harding}, {Hasselquist}, {Hayes}, {Hearty}, {Hern{\'a}ndez}, {Hill}, {Hogg}, {Holtzman}, {Horta}, {Hsieh}, {Hsu}, {Hsu}, {Huber}, {Huertas-Company}, {Hutchinson}, {Hwang}, {Ibarra-Medel}, {Chitham}, {Ilha}, {Imig}, {Jaekle}, {Jayasinghe}, {Ji}, {Johnson}, {Jones}, {J{\"o}nsson}, {Katkov}, {Khalatyan}, {Kinemuchi}, {Kisku}, {Knapen}, {Kneib}, {Kollmeier}, {Kong}, {Kounkel}, {Kreckel}, {Krishnarao}, {Lacerna}, {Lane}, {Langgin}, {Lavender}, {Law}, {Lazarz}, {Leung}, {Leung}, {Lewis}, {Li}, {Li}, {Lian}, {Liang}, {Lin}, {Lin}, {Lin}, {Lintott}, {Long}, {Longa-Pe{\~n}a}, {L{\'o}pez-Cob{\'a}}, {Lu},
  {Lundgren}, {Luo}, {Mackereth}, {de la Macorra}, {Mahadevan}, {Majewski}, {Manchado}, {Mandeville}, {Maraston}, {Margalef-Bentabol}, {Masseron}, {Masters}, {Mathur}, {McDermid}, {Mckay}, {Merloni}, {Merrifield}, {Meszaros}, {Miglio}, {Di Mille}, {Minniti}, {Minsley}, {Monachesi}, {Moon}, {Mosser}, {Mulchaey}, {Muna}, {Mu{\~n}oz}, {Myers}, {Myers}, {Nadathur}, {Nair}, {Nandra}, {Neumann}, {Newman}, {Nidever}, {Nikakhtar}, {Nitschelm}, {O'Connell}, {Garma-Oehmichen}, {Luan Souza de Oliveira}, {Olney}, {Oravetz}, {Ortigoza-Urdaneta}, {Osorio}, {Otter}, {Pace}, {Padilla}, {Pan}, {Pan}, {Parikh}, {Parker}, {Peirani}, {Pe{\~n}a Ram{\'\i}rez}, {Penny}, {Percival}, {Perez-Fournon}, {Pinsonneault}, {Poidevin}, {Poovelil}, {Price-Whelan}, {B{\'a}rbara de Andrade Queiroz}, {Raddick}, {Ray}, {Rembold}, {Riddle}, {Riffel}, {Riffel}, {Rix}, {Robin}, {Rodr{\'\i}guez-Puebla}, {Roman-Lopes}, {Rom{\'a}n-Z{\'u}{\~n}iga}, {Rose}, {Ross}, {Rossi}, {Rubin}, {Salvato}, {S{\'a}nchez}, {S{\'a}nchez-Gallego}, {Sanderson}, {Santana
  Rojas}, {Sarceno}, {Sarmiento}, {Sayres}, {Sazonova}, {Schaefer}, {Schiavon}, {Schlegel}, {Schneider}, {Schultheis}, {Schwope}, {Serenelli}, {Serna}, {Shao}, {Shapiro}, {Sharma}, {Shen}, {Shetrone}, {Shu}, {Simon}, {Skrutskie}, {Smethurst}, {Smith}, {Sobeck}, {Spoo}, {Sprague}, {Stark}, {Stassun}, {Steinmetz}, {Stello}, {Stone-Martinez}, {Storchi-Bergmann}, {Stringfellow}, {Stutz}, {Su}, {Taghizadeh-Popp}, {Talbot}, {Tayar}, {Telles}, {Teske}, {Thakar}, {Theissen}, {Tkachenko}, {Thomas}, {Tojeiro}, {Hernandez Toledo}, {Troup}, {Trump}, {Trussler}, {Turner}, {Tuttle}, {Unda-Sanzana}, {V{\'a}zquez-Mata}, {Valentini}, {Valenzuela}, {Vargas-Gonz{\'a}lez}, {Vargas-Maga{\~n}a}, {Alfaro}, {Villanova}, {Vincenzo}, {Wake}, {Warfield}, {Washington}, {Weaver}, {Weijmans}, {Weinberg}, {Weiss}, {Westfall}, {Wild}, {Wilde}, {Wilson}, {Wilson}, {Wilson}, {Wolf}, {Wood-Vasey}, {Yan}, {Zamora}, {Zasowski}, {Zhang}, {Zhao}, {Zheng}, {Zheng}, \& {Zhu}}]{apogeeDR17}
{Abdurro'uf}, {Accetta}, K., {Aerts}, C., {et~al.} 2022, \apjs, 259, 35

\bibitem[{{Agertz} {et~al.}(2021){Agertz}, {Renaud}, {Feltzing}, {Read}, {Ryde}, {Andersson}, {Rey}, {Bensby}, \& {Feuillet}}]{Agertz2021_vintergatanI}
{Agertz}, O., {Renaud}, F., {Feltzing}, S., {et~al.} 2021, \mnras, 503, 5826

\bibitem[{{Anders} {et~al.}(2023){Anders}, {Gispert}, {Ratcliffe}, {Chiappini}, {Minchev}, {Nepal}, {Queiroz}, {Amarante}, {Antoja}, {Casali}, {Casamiquela}, {Khalatyan}, {Miglio}, {Perottoni}, \& {Schultheis}}]{Anders2023_ages}
{Anders}, F., {Gispert}, P., {Ratcliffe}, B., {et~al.} 2023, \aap, 678, A158

\bibitem[{{Annem} \& {Khoperskov}(2024)}]{Annem2024}
{Annem}, B. \& {Khoperskov}, S. 2024, \mnras, 527, 2426

\bibitem[{{Arellano-C{\'o}rdova} {et~al.}(2021){Arellano-C{\'o}rdova}, {Esteban}, {Garc{\'\i}a-Rojas}, \& {M{\'e}ndez-Delgado}}]{ArellanoCordova2021}
{Arellano-C{\'o}rdova}, K.~Z., {Esteban}, C., {Garc{\'\i}a-Rojas}, J., \& {M{\'e}ndez-Delgado}, J.~E. 2021, \mnras, 502, 225

\bibitem[{{Asplund} {et~al.}(2009){Asplund}, {Grevesse}, {Sauval}, \& {Scott}}]{Asplund2009}
{Asplund}, M., {Grevesse}, N., {Sauval}, A.~J., \& {Scott}, P. 2009, \araa, 47, 481

\bibitem[{{Bellardini} {et~al.}(2022){Bellardini}, {Wetzel}, {Loebman}, \& {Bailin}}]{Bellardini2022}
{Bellardini}, M.~A., {Wetzel}, A., {Loebman}, S.~R., \& {Bailin}, J. 2022, \mnras, 514, 4270

\bibitem[{{Belokurov} {et~al.}(2018){Belokurov}, {Erkal}, {Evans}, {Koposov}, \& {Deason}}]{Belokurov2018}
{Belokurov}, V., {Erkal}, D., {Evans}, N.~W., {Koposov}, S.~E., \& {Deason}, A.~J. 2018, \mnras, 478, 611

\bibitem[{{Bland-Hawthorn} \& {Gerhard}(2016)}]{BH2016}
{Bland-Hawthorn}, J. \& {Gerhard}, O. 2016, \araa, 54, 529

\bibitem[{{Bland-Hawthorn} {et~al.}(2010){Bland-Hawthorn}, {Krumholz}, \& {Freeman}}]{BH2010}
{Bland-Hawthorn}, J., {Krumholz}, M.~R., \& {Freeman}, K. 2010, \apj, 713, 166

\bibitem[{{Buck}(2020)}]{2020_buckchemical}
{Buck}, T. 2020, \mnras, 491, 5435

\bibitem[{{Buck} {et~al.}(2020){Buck}, {Obreja}, {Macci{\`o}}, {Minchev}, {Dutton}, \& {Ostriker}}]{2020buck_NIHAO-UHD}
{Buck}, T., {Obreja}, A., {Macci{\`o}}, A.~V., {et~al.} 2020, \mnras, 491, 3461

\bibitem[{{Buck} {et~al.}(2023){Buck}, {Obreja}, {Ratcliffe}, {Lu}, {Minchev}, \& {Macci{\`o}}}]{Buck2023}
{Buck}, T., {Obreja}, A., {Ratcliffe}, B., {et~al.} 2023, \mnras, 523, 1565

\bibitem[{{Buck} {et~al.}(2021){Buck}, {Rybizki}, {Buder}, {Obreja}, {Macci{\`o}}, {Pfrommer}, {Steinmetz}, \& {Ness}}]{2021BuckHD_chemEnrich}
{Buck}, T., {Rybizki}, J., {Buder}, S., {et~al.} 2021, \mnras, 508, 3365

\bibitem[{{Carr} {et~al.}(2022){Carr}, {Johnston}, {Laporte}, \& {Ness}}]{Carr2022}
{Carr}, C., {Johnston}, K.~V., {Laporte}, C. F.~P., \& {Ness}, M.~K. 2022, \mnras, 516, 5067

\bibitem[{{Carrillo} {et~al.}(2023){Carrillo}, {Ness}, {Hawkins}, {Sanderson}, {Wang}, {Wetzel}, \& {Bellardini}}]{Carrillo2023}
{Carrillo}, A., {Ness}, M.~K., {Hawkins}, K., {et~al.} 2023, \apj, 942, 35

\bibitem[{{Casamiquela} {et~al.}(2021){Casamiquela}, {Castro-Ginard}, {Anders}, \& {Soubiran}}]{Casamiquela2021}
{Casamiquela}, L., {Castro-Ginard}, A., {Anders}, F., \& {Soubiran}, C. 2021, \aap, 654, A151

\bibitem[{{Deharveng} {et~al.}(2000){Deharveng}, {Pe{\~n}a}, {Caplan}, \& {Costero}}]{Deharveng2000}
{Deharveng}, L., {Pe{\~n}a}, M., {Caplan}, J., \& {Costero}, R. 2000, \mnras, 311, 329

\bibitem[{{Di Matteo} {et~al.}(2013){Di Matteo}, {Haywood}, {Combes}, {Semelin}, \& {Snaith}}]{DiMatteo2013}
{Di Matteo}, P., {Haywood}, M., {Combes}, F., {Semelin}, B., \& {Snaith}, O.~N. 2013, \aap, 553, A102

\bibitem[{{Doherty} {et~al.}(2014){Doherty}, {Gil-Pons}, {Lau}, {Lattanzio}, {Siess}, \& {Campbell}}]{Doherty2014}
{Doherty}, C.~L., {Gil-Pons}, P., {Lau}, H. H.~B., {et~al.} 2014, \mnras, 441, 582

\bibitem[{{El-Badry} {et~al.}(2016){El-Badry}, {Wetzel}, {Geha}, {Hopkins}, {Kere{\v{s}}}, {Chan}, \& {Faucher-Gigu{\`e}re}}]{ElBadry2016}
{El-Badry}, K., {Wetzel}, A., {Geha}, M., {et~al.} 2016, \apj, 820, 131

\bibitem[{{Esteban} {et~al.}(2017){Esteban}, {Fang}, {Garc{\'\i}a-Rojas}, \& {Toribio San Cipriano}}]{Esteban2017}
{Esteban}, C., {Fang}, X., {Garc{\'\i}a-Rojas}, J., \& {Toribio San Cipriano}, L. 2017, \mnras, 471, 987

\bibitem[{{Feltzing} {et~al.}(2020){Feltzing}, {Bowers}, \& {Agertz}}]{2020Feltzing}
{Feltzing}, S., {Bowers}, J.~B., \& {Agertz}, O. 2020, \mnras, 493, 1419

\bibitem[{{Fishlock} {et~al.}(2014){Fishlock}, {Karakas}, {Lugaro}, \& {Yong}}]{Fishlock2014}
{Fishlock}, C.~K., {Karakas}, A.~I., {Lugaro}, M., \& {Yong}, D. 2014, \apj, 797, 44

\bibitem[{{Frankel} {et~al.}(2018){Frankel}, {Rix}, {Ting}, {Ness}, \& {Hogg}}]{Frankel2018}
{Frankel}, N., {Rix}, H.-W., {Ting}, Y.-S., {Ness}, M., \& {Hogg}, D.~W. 2018, \apj, 865, 96

\bibitem[{{Frankel} {et~al.}(2019){Frankel}, {Sanders}, {Rix}, {Ting}, \& {Ness}}]{2019Frankel}
{Frankel}, N., {Sanders}, J., {Rix}, H.-W., {Ting}, Y.-S., \& {Ness}, M. 2019, \apj, 884, 99

\bibitem[{{Frankel} {et~al.}(2020){Frankel}, {Sanders}, {Ting}, \& {Rix}}]{Frankel2020}
{Frankel}, N., {Sanders}, J., {Ting}, Y.-S., \& {Rix}, H.-W. 2020, \apj, 896, 15

\bibitem[{{Freeman} \& {Bland-Hawthorn}(2002)}]{2002freeman-BH}
{Freeman}, K. \& {Bland-Hawthorn}, J. 2002, \araa, 40, 487

\bibitem[{{Garrison-Kimmel} {et~al.}(2018){Garrison-Kimmel}, {Hopkins}, {Wetzel}, {El-Badry}, {Sanderson}, {Bullock}, {Ma}, {van de Voort}, {Hafen}, {Faucher-Gigu{\`e}re}, {Hayward}, {Quataert}, {Kere{\v{s}}}, \& {Boylan-Kolchin}}]{2018FIRE2}
{Garrison-Kimmel}, S., {Hopkins}, P.~F., {Wetzel}, A., {et~al.} 2018, \mnras, 481, 4133

\bibitem[{{G{\'e}ron} {et~al.}(2024){G{\'e}ron}, {Smethurst}, {Lintott}, {Masters}, {Garland}, {Mengistu}, {O'Ryan}, \& {Simmons}}]{Geron2024}
{G{\'e}ron}, T., {Smethurst}, R.~J., {Lintott}, C., {et~al.} 2024, arXiv e-prints, arXiv:2405.05960

\bibitem[{{Grand} \& {Kawata}(2016)}]{Grand2016_radialMigration}
{Grand}, R.~J.~J. \& {Kawata}, D. 2016, Astronomische Nachrichten, 337, 957

\bibitem[{{Haywood} {et~al.}(2018){Haywood}, {Di Matteo}, {Lehnert}, {Snaith}, {Khoperskov}, \& {G{\'o}mez}}]{Haywood2018}
{Haywood}, M., {Di Matteo}, P., {Lehnert}, M.~D., {et~al.} 2018, \apj, 863, 113

\bibitem[{{Helmi} {et~al.}(2018){Helmi}, {Babusiaux}, {Koppelman}, {Massari}, {Veljanoski}, \& {Brown}}]{Helmi2018_gse}
{Helmi}, A., {Babusiaux}, C., {Koppelman}, H.~H., {et~al.} 2018, \nat, 563, 85

\bibitem[{{Ibata} {et~al.}(1994){Ibata}, {Gilmore}, \& {Irwin}}]{Ibata1994}
{Ibata}, R.~A., {Gilmore}, G., \& {Irwin}, M.~J. 1994, \nat, 370, 194

\bibitem[{{J{\"o}nsson} {et~al.}(2020){J{\"o}nsson}, {Holtzman}, {Allende Prieto}, {Cunha}, {Garc{\'\i}a-Hern{\'a}ndez}, {Hasselquist}, {Masseron}, {Osorio}, {Shetrone}, {Smith}, {Stringfellow}, {Bizyaev}, {Edvardsson}, {Majewski}, {M{\'e}sz{\'a}ros}, {Souto}, {Zamora}, {Beaton}, {Bovy}, {Donor}, {Pinsonneault}, {Poovelil}, \& {Sobeck}}]{Jonsson2020}
{J{\"o}nsson}, H., {Holtzman}, J.~A., {Allende Prieto}, C., {et~al.} 2020, \aj, 160, 120

\bibitem[{{Karakas}(2010)}]{Karakas2010}
{Karakas}, A.~I. 2010, \mnras, 403, 1413

\bibitem[{{Khoperskov} {et~al.}(2020){Khoperskov}, {Di Matteo}, {Haywood}, {G{\'o}mez}, \& {Snaith}}]{Khoperskov_bar_migration2018}
{Khoperskov}, S., {Di Matteo}, P., {Haywood}, M., {G{\'o}mez}, A., \& {Snaith}, O.~N. 2020, \aap, 638, A144

\bibitem[{{Khoperskov} {et~al.}(2018){Khoperskov}, {Haywood}, {Di Matteo}, {Lehnert}, \& {Combes}}]{Khoperskov2018}
{Khoperskov}, S., {Haywood}, M., {Di Matteo}, P., {Lehnert}, M.~D., \& {Combes}, F. 2018, \aap, 609, A60

\bibitem[{{Khoperskov} {et~al.}(2023){Khoperskov}, {Minchev}, {Libeskind}, {Haywood}, {Di Matteo}, {Belokurov}, {Steinmetz}, {Gomez}, {Grand}, {Hoffman}, {Knebe}, {Sorce}, {Spaare}, {Tempel}, \& {Vogelsberger}}]{Khoperskov2023_Hestia}
{Khoperskov}, S., {Minchev}, I., {Libeskind}, N., {et~al.} 2023, \aap, 677, A89

\bibitem[{{Khoperskov} {et~al.}(2024){Khoperskov}, {Minchev}, {Steinmetz}, {Ratcliffe}, {Walcher}, \& {Libeskind}}]{Khoperskov2023}
{Khoperskov}, S., {Minchev}, I., {Steinmetz}, M., {et~al.} 2024, \mnras, 533, 3975

\bibitem[{{Kobayashi} {et~al.}(2006){Kobayashi}, {Umeda}, {Nomoto}, {Tominaga}, \& {Ohkubo}}]{Kobayashi2006}
{Kobayashi}, C., {Umeda}, H., {Nomoto}, K., {Tominaga}, N., \& {Ohkubo}, T. 2006, \apj, 653, 1145

\bibitem[{{Kubryk} {et~al.}(2013){Kubryk}, {Prantzos}, \& {Athanassoula}}]{Kubryk2013}
{Kubryk}, M., {Prantzos}, N., \& {Athanassoula}, E. 2013, \mnras, 436, 1479

\bibitem[{{Law} \& {Majewski}(2010)}]{Law2010}
{Law}, D.~R. \& {Majewski}, S.~R. 2010, \apj, 714, 229

\bibitem[{{Lian} {et~al.}(2023){Lian}, {Bergemann}, {Pillepich}, {Zasowski}, \& {Lane}}]{Lian2023}
{Lian}, J., {Bergemann}, M., {Pillepich}, A., {Zasowski}, G., \& {Lane}, R.~R. 2023, Nature Astronomy, 7, 951

\bibitem[{{Libeskind} {et~al.}(2020){Libeskind}, {Carlesi}, {Grand}, {Khalatyan}, {Knebe}, {Pakmor}, {Pilipenko}, {Pawlowski}, {Sparre}, {Tempel}, {Wang}, {Courtois}, {Gottl{\"o}ber}, {Hoffman}, {Minchev}, {Pfrommer}, {Sorce}, {Springel}, {Steinmetz}, {Tully}, {Vogelsberger}, \& {Yepes}}]{Libeskind2020}
{Libeskind}, N.~I., {Carlesi}, E., {Grand}, R. J.~J., {et~al.} 2020, \mnras, 498, 2968

\bibitem[{{Lindegren} \& {Feltzing}(2013)}]{2013Lindegren}
{Lindegren}, L. \& {Feltzing}, S. 2013, \aap, 553, A94

\bibitem[{{Lu} {et~al.}(2022{\natexlab{a}}){Lu}, {Buck}, {Minchev}, \& {Ness}}]{Lu2022_sims}
{Lu}, Y., {Buck}, T., {Minchev}, I., \& {Ness}, M.~K. 2022{\natexlab{a}}, \mnras, 515, L34

\bibitem[{{Lu} {et~al.}(2024){Lu}, {Buck}, {Nidever}, {Ratcliffe}, {Minchev}, {Macci{\`o}}, \& {Obreja}}]{Lu2024_RbLMC}
{Lu}, Y., {Buck}, T., {Nidever}, D., {et~al.} 2024, \mnras, 532, 411

\bibitem[{{Lu} {et~al.}(2022{\natexlab{b}}){Lu}, {Minchev}, {Buck}, {Khoperskov}, {Steinmetz}, {Libeskind}, {Cescutti}, {Freeman}, \& {Ratcliffe}}]{Lu2022_Rb}
{Lu}, Y., {Minchev}, I., {Buck}, T., {et~al.} 2022{\natexlab{b}}, arXiv e-prints, arXiv:2212.04515

\bibitem[{{Majewski} {et~al.}(2017){Majewski}, {Schiavon}, {Frinchaboy}, {Allende Prieto}, {Barkhouser}, {Bizyaev}, {Blank}, {Brunner}, {Burton}, {Carrera}, {Chojnowski}, {Cunha}, {Epstein}, {Fitzgerald}, {Garc{\'{\i}}a P{\'e}rez}, {Hearty}, {Henderson}, {Holtzman}, {Johnson}, {Lam}, {Lawler}, {Maseman}, {M{\'e}sz{\'a}ros}, {Nelson}, {Nguyen}, {Nidever}, {Pinsonneault}, {Shetrone}, {Smee}, {Smith}, {Stolberg}, {Skrutskie}, {Walker}, {Wilson}, {Zasowski}, {Anders}, {Basu}, {Beland}, {Blanton}, {Bovy}, {Brownstein}, {Carlberg}, {Chaplin}, {Chiappini}, {Eisenstein}, {Elsworth}, {Feuillet}, {Fleming}, {Galbraith-Frew}, {Garc{\'{\i}}a}, {Garc{\'{\i}}a-Hern{\'a}ndez}, {Gillespie}, {Girardi}, {Gunn}, {Hasselquist}, {Hayden}, {Hekker}, {Ivans}, {Kinemuchi}, {Klaene}, {Mahadevan}, {Mathur}, {Mosser}, {Muna}, {Munn}, {Nichol}, {O'Connell}, {Parejko}, {Robin}, {Rocha-Pinto}, {Schultheis}, {Serenelli}, {Shane}, {Silva Aguirre}, {Sobeck}, {Thompson}, {Troup}, {Weinberg}, \& {Zamora}}]{Majewski2017}
{Majewski}, S.~R., {Schiavon}, R.~P., {Frinchaboy}, P.~M., {et~al.} 2017, \aj, 154, 94

\bibitem[{{Marinacci} {et~al.}(2018){Marinacci}, {Vogelsberger}, {Pakmor}, {Torrey}, {Springel}, {Hernquist}, {Nelson}, {Weinberger}, {Pillepich}, {Naiman}, \& {Genel}}]{Marinacci2018_TNG}
{Marinacci}, F., {Vogelsberger}, M., {Pakmor}, R., {et~al.} 2018, \mnras, 480, 5113

\bibitem[{{Martig} {et~al.}(2009){Martig}, {Bournaud}, {Teyssier}, \& {Dekel}}]{2009ApJ...707..250M}
{Martig}, M., {Bournaud}, F., {Teyssier}, R., \& {Dekel}, A. 2009, \apj, 707, 250

\bibitem[{{Matteucci} \& {Francois}(1989)}]{Matteucci1989}
{Matteucci}, F. \& {Francois}, P. 1989, \mnras, 239, 885

\bibitem[{{Minchev} {et~al.}(2018){Minchev}, {Anders}, {Recio-Blanco}, {Chiappini}, {de Laverny}, {Queiroz}, {Steinmetz}, {Adibekyan}, {Carrillo}, {Cescutti}, {Guiglion}, {Hayden}, {de Jong}, {Kordopatis}, {Majewski}, {Martig}, \& {Santiago}}]{Minchev2018_rbirth}
{Minchev}, I., {Anders}, F., {Recio-Blanco}, A., {et~al.} 2018, \mnras, 481, 1645

\bibitem[{{Minchev} {et~al.}(2013){Minchev}, {Chiappini}, \& {Martig}}]{Minchev2013}
{Minchev}, I., {Chiappini}, C., \& {Martig}, M. 2013, \aap, 558, A9

\bibitem[{{Minchev} {et~al.}(2014){Minchev}, {Chiappini}, \& {Martig}}]{Minchev2014}
{Minchev}, I., {Chiappini}, C., \& {Martig}, M. 2014, \aap, 572, A92

\bibitem[{{Minchev} \& {Famaey}(2010)}]{Minchev2010}
{Minchev}, I. \& {Famaey}, B. 2010, \apj, 722, 112

\bibitem[{{Minchev} {et~al.}(2012{\natexlab{a}}){Minchev}, {Famaey}, {Quillen}, \& {Dehnen}}]{Minchev2012a}
{Minchev}, I., {Famaey}, B., {Quillen}, A.~C., \& {Dehnen}, W. 2012{\natexlab{a}}, in European Physical Journal Web of Conferences, Vol.~19, European Physical Journal Web of Conferences, 07002

\bibitem[{{Minchev} {et~al.}(2012{\natexlab{b}}){Minchev}, {Famaey}, {Quillen}, {Dehnen}, {Martig}, \& {Siebert}}]{Minchev2012}
{Minchev}, I., {Famaey}, B., {Quillen}, A.~C., {et~al.} 2012{\natexlab{b}}, \aap, 548, A127

\bibitem[{{Minchev} \& {Quillen}(2006)}]{Minchev2006}
{Minchev}, I. \& {Quillen}, A.~C. 2006, \mnras, 368, 623

\bibitem[{{Moll{\'a}} {et~al.}(2019){Moll{\'a}}, {D{\'\i}az}, {Cavichia}, {Gibson}, {Maciel}, {Costa}, {Ascasibar}, \& {Few}}]{Molla2019}
{Moll{\'a}}, M., {D{\'\i}az}, {\'A}.~I., {Cavichia}, O., {et~al.} 2019, \mnras, 482, 3071

\bibitem[{{Naiman} {et~al.}(2018){Naiman}, {Pillepich}, {Springel}, {Ramirez-Ruiz}, {Torrey}, {Vogelsberger}, {Pakmor}, {Nelson}, {Marinacci}, {Hernquist}, {Weinberger}, \& {Genel}}]{Naiman2018_TNG}
{Naiman}, J.~P., {Pillepich}, A., {Springel}, V., {et~al.} 2018, \mnras, 477, 1206

\bibitem[{{Nelson} {et~al.}(2019{\natexlab{a}}){Nelson}, {Pillepich}, {Springel}, {Pakmor}, {Weinberger}, {Genel}, {Torrey}, {Vogelsberger}, {Marinacci}, \& {Hernquist}}]{Nelson2019}
{Nelson}, D., {Pillepich}, A., {Springel}, V., {et~al.} 2019{\natexlab{a}}, \mnras, 490, 3234

\bibitem[{{Nelson} {et~al.}(2019{\natexlab{b}}){Nelson}, {Pillepich}, {Springel}, {Pakmor}, {Weinberger}, {Genel}, {Torrey}, {Vogelsberger}, {Marinacci}, \& {Hernquist}}]{2019MNRAS.490.3234N}
{Nelson}, D., {Pillepich}, A., {Springel}, V., {et~al.} 2019{\natexlab{b}}, \mnras, 490, 3234

\bibitem[{{Nelson} {et~al.}(2018{\natexlab{a}}){Nelson}, {Pillepich}, {Springel}, {Weinberger}, {Hernquist}, {Pakmor}, {Genel}, {Torrey}, {Vogelsberger}, {Kauffmann}, {Marinacci}, \& {Naiman}}]{Nelson2018_TNG}
{Nelson}, D., {Pillepich}, A., {Springel}, V., {et~al.} 2018{\natexlab{a}}, \mnras, 475, 624

\bibitem[{{Nelson} {et~al.}(2018{\natexlab{b}}){Nelson}, {Pillepich}, {Springel}, {Weinberger}, {Hernquist}, {Pakmor}, {Genel}, {Torrey}, {Vogelsberger}, {Kauffmann}, {Marinacci}, \& {Naiman}}]{2018MNRAS.475..624N}
{Nelson}, D., {Pillepich}, A., {Springel}, V., {et~al.} 2018{\natexlab{b}}, \mnras, 475, 624

\bibitem[{{Nelson} {et~al.}(2019{\natexlab{c}}){Nelson}, {Springel}, {Pillepich}, {Rodriguez-Gomez}, {Torrey}, {Genel}, {Vogelsberger}, {Pakmor}, {Marinacci}, {Weinberger}, {Kelley}, {Lovell}, {Diemer}, \& {Hernquist}}]{Nelson2019a_TNG}
{Nelson}, D., {Springel}, V., {Pillepich}, A., {et~al.} 2019{\natexlab{c}}, Computational Astrophysics and Cosmology, 6, 2

\bibitem[{{Netopil} {et~al.}(2022){Netopil}, {Oralhan}, {{\c{C}}akmak}, {Michel}, \& {Karata{\c{s}}}}]{Netopil2022}
{Netopil}, M., {Oralhan}, {\.I}.~A., {{\c{C}}akmak}, H., {Michel}, R., \& {Karata{\c{s}}}, Y. 2022, \mnras, 509, 421

\bibitem[{{Nomoto} {et~al.}(1997){Nomoto}, {Iwamoto}, {Nakasato}, {Thielemann}, {Brachwitz}, {Tsujimoto}, {Kubo}, \& {Kishimoto}}]{Nomoto1997}
{Nomoto}, K., {Iwamoto}, K., {Nakasato}, N., {et~al.} 1997, \nphysa, 621, 467

\bibitem[{{Pessa} {et~al.}(2023){Pessa}, {Schinnerer}, {Sanchez-Blazquez}, {Belfiore}, {Groves}, {Emsellem}, {Neumann}, {Leroy}, {Bigiel}, {Chevance}, {Dale}, {Glover}, {Grasha}, {Klessen}, {Kreckel}, {Kruijssen}, {Pinna}, {Querejeta}, {Rosolowsky}, \& {Williams}}]{2023A&A...673A.147P}
{Pessa}, I., {Schinnerer}, E., {Sanchez-Blazquez}, P., {et~al.} 2023, \aap, 673, A147

\bibitem[{{Pilkington} {et~al.}(2012){Pilkington}, {Few}, {Gibson}, {Calura}, {Michel-Dansac}, {Thacker}, {Moll{\'a}}, {Matteucci}, {Rahimi}, {Kawata}, {Kobayashi}, {Brook}, {Stinson}, {Couchman}, {Bailin}, \& {Wadsley}}]{Pilkington2012}
{Pilkington}, K., {Few}, C.~G., {Gibson}, B.~K., {et~al.} 2012, \aap, 540, A56

\bibitem[{{Pillepich} {et~al.}(2018){Pillepich}, {Nelson}, {Hernquist}, {Springel}, {Pakmor}, {Torrey}, {Weinberger}, {Genel}, {Naiman}, {Marinacci}, \& {Vogelsberger}}]{Pillepich2018_TNG}
{Pillepich}, A., {Nelson}, D., {Hernquist}, L., {et~al.} 2018, \mnras, 475, 648

\bibitem[{{Pillepich} {et~al.}(2019){Pillepich}, {Nelson}, {Springel}, {Pakmor}, {Torrey}, {Weinberger}, {Vogelsberger}, {Marinacci}, {Genel}, {van der Wel}, \& {Hernquist}}]{Pillepich2019}
{Pillepich}, A., {Nelson}, D., {Springel}, V., {et~al.} 2019, \mnras, 490, 3196

\bibitem[{{Pillepich} {et~al.}(2023){Pillepich}, {Sotillo-Ramos}, {Ramesh}, {Nelson}, {Engler}, {Rodriguez-Gomez}, {Fournier}, {Donnari}, {Springel}, \& {Hernquist}}]{Pillepich2023_MW_Sample}
{Pillepich}, A., {Sotillo-Ramos}, D., {Ramesh}, R., {et~al.} 2023, arXiv e-prints, arXiv:2303.16217

\bibitem[{{Portinari} {et~al.}(1998){Portinari}, {Chiosi}, \& {Bressan}}]{Portinari1998}
{Portinari}, L., {Chiosi}, C., \& {Bressan}, A. 1998, \aap, 334, 505

\bibitem[{{Prantzos} {et~al.}(2023){Prantzos}, {Abia}, {Chen}, {de Laverny}, {Recio-Blanco}, {Athanassoula}, {Roberti}, {Vescovi}, {Limongi}, {Chieffi}, \& {Cristallo}}]{Prantzos2023}
{Prantzos}, N., {Abia}, C., {Chen}, T., {et~al.} 2023, \mnras, 523, 2126

\bibitem[{{Queiroz} {et~al.}(2023){Queiroz}, {Anders}, {Chiappini}, {Khalatyan}, {Santiago}, {Nepal}, {Steinmetz}, {Gallart}, {Valentini}, {Dal Ponte}, {Barbuy}, {P{\'e}rez-Villegas}, {Masseron}, {Fern{\'a}ndez-Trincado}, {Khoperskov}, {Minchev}, {Fern{\'a}ndez-Alvar}, {Lane}, \& {Nitschelm}}]{Queiroz2023_SH}
{Queiroz}, A.~B.~A., {Anders}, F., {Chiappini}, C., {et~al.} 2023, \aap, 673, A155

\bibitem[{{Querejeta} {et~al.}(2021){Querejeta}, {Schinnerer}, {Meidt}, {Sun}, {Leroy}, {Emsellem}, {Klessen}, {Mu{\~n}oz-Mateos}, {Salo}, {Laurikainen}, {Be{\v{s}}li{\'c}}, {Blanc}, {Chevance}, {Dale}, {Eibensteiner}, {Faesi}, {Garc{\'\i}a-Rodr{\'\i}guez}, {Glover}, {Grasha}, {Henshaw}, {Herrera}, {Hughes}, {Kreckel}, {Kruijssen}, {Liu}, {Murphy}, {Pan}, {Pety}, {Razza}, {Rosolowsky}, {Saito}, {Schruba}, {Usero}, {Watkins}, \& {Williams}}]{2021A&A...656A.133Q}
{Querejeta}, M., {Schinnerer}, E., {Meidt}, S., {et~al.} 2021, \aap, 656, A133

\bibitem[{{Quillen} {et~al.}(2009){Quillen}, {Minchev}, {Bland-Hawthorn}, \& {Haywood}}]{Quillen2009}
{Quillen}, A.~C., {Minchev}, I., {Bland-Hawthorn}, J., \& {Haywood}, M. 2009, \mnras, 397, 1599

\bibitem[{{Ratcliffe} {et~al.}(2023){Ratcliffe}, {Minchev}, {Anders}, {Khoperskov}, {Guiglion}, {Buck}, {Cunha}, {Queiroz}, {Nitschelm}, {Meszaros}, {Steinmetz}, {de Jong}, {Nepal}, {Lane}, \& {Sobeck}}]{Ratcliffe2023_enrichment}
{Ratcliffe}, B., {Minchev}, I., {Anders}, F., {et~al.} 2023, \mnras, 525, 2208

\bibitem[{{Ratcliffe} {et~al.}(2024){Ratcliffe}, {Minchev}, {Cescutti}, {Spitoni}, {J{\"o}nsson}, {Anders}, {Queiroz}, \& {Steinmetz}}]{Ratcliffe2023_chemicalclocks}
{Ratcliffe}, B., {Minchev}, I., {Cescutti}, G., {et~al.} 2024, \mnras, 528, 3464

\bibitem[{{Ratcliffe} {et~al.}(2022){Ratcliffe}, {Ness}, {Buck}, {Johnston}, {Sen}, {Beraldo e Silva}, \& {Debattista}}]{2022Ratcliffe}
{Ratcliffe}, B.~L., {Ness}, M.~K., {Buck}, T., {et~al.} 2022, \apj, 924, 60

\bibitem[{{Ratcliffe} {et~al.}(2020){Ratcliffe}, {Ness}, {Johnston}, \& {Sen}}]{2020Ratcliffe}
{Ratcliffe}, B.~L., {Ness}, M.~K., {Johnston}, K.~V., \& {Sen}, B. 2020, \apj, 900, 165

\bibitem[{{Renaud} {et~al.}(2021{\natexlab{a}}){Renaud}, {Agertz}, {Andersson}, {Read}, {Ryde}, {Bensby}, {Rey}, \& {Feuillet}}]{Renaud2021_vintergatanIII}
{Renaud}, F., {Agertz}, O., {Andersson}, E.~P., {et~al.} 2021{\natexlab{a}}, \mnras, 503, 5868

\bibitem[{{Renaud} {et~al.}(2021{\natexlab{b}}){Renaud}, {Agertz}, {Read}, {Ryde}, {Andersson}, {Bensby}, {Rey}, \& {Feuillet}}]{Renaud2021_vintergatanII}
{Renaud}, F., {Agertz}, O., {Read}, J.~I., {et~al.} 2021{\natexlab{b}}, \mnras, 503, 5846

\bibitem[{{Renaud} {et~al.}(2019){Renaud}, {Bournaud}, {Agertz}, {Kraljic}, {Schinnerer}, {Bolatto}, {Daddi}, \& {Hughes}}]{2019A&A...625A..65R}
{Renaud}, F., {Bournaud}, F., {Agertz}, O., {et~al.} 2019, \aap, 625, A65

\bibitem[{{Renaud} {et~al.}(2015){Renaud}, {Bournaud}, {Emsellem}, {Agertz}, {Athanassoula}, {Combes}, {Elmegreen}, {Kraljic}, {Motte}, \& {Teyssier}}]{2015MNRAS.454.3299R}
{Renaud}, F., {Bournaud}, F., {Emsellem}, E., {et~al.} 2015, \mnras, 454, 3299

\bibitem[{{Renaud} {et~al.}(2024){Renaud}, {Ratcliffe}, {Minchev}, {Haywood}, {Di Matteo}, {Agertz}, \& {Romeo}}]{Renaud2024}
{Renaud}, F., {Ratcliffe}, B., {Minchev}, I., {et~al.} 2024, arXiv e-prints, arXiv:2409.10598

\bibitem[{{Rodriguez-Gomez} {et~al.}(2015){Rodriguez-Gomez}, {Genel}, {Vogelsberger}, {Sijacki}, {Pillepich}, {Sales}, {Torrey}, {Snyder}, {Nelson}, {Springel}, {Ma}, \& {Hernquist}}]{RodriguezGomez2015}
{Rodriguez-Gomez}, V., {Genel}, S., {Vogelsberger}, M., {et~al.} 2015, \mnras, 449, 49

\bibitem[{{Ro{\v s}kar} {et~al.}(2008){Ro{\v s}kar}, {Debattista}, {Quinn}, {Stinson}, \& {Wadsley}}]{Roskar2008_migration}
{Ro{\v s}kar}, R., {Debattista}, V.~P., {Quinn}, T.~R., {Stinson}, G.~S., \& {Wadsley}, J. 2008, \apjl, 684, L79

\bibitem[{{Sch{\"o}nrich} \& {Binney}(2009)}]{2009schonrichBinney}
{Sch{\"o}nrich}, R. \& {Binney}, J. 2009, \mnras, 399, 1145

\bibitem[{{Sellwood} \& {Binney}(2002)}]{Selwood2002}
{Sellwood}, J.~A. \& {Binney}, J.~J. 2002, \mnras, 336, 785

\bibitem[{{Sotillo-Ramos} {et~al.}(2022){Sotillo-Ramos}, {Pillepich}, {Donnari}, {Nelson}, {Eisert}, {Rodriguez-Gomez}, {Joshi}, {Vogelsberger}, \& {Hernquist}}]{SotilloRamos2022}
{Sotillo-Ramos}, D., {Pillepich}, A., {Donnari}, M., {et~al.} 2022, \mnras, 516, 5404

\bibitem[{{Springel}(2010)}]{Springel2010}
{Springel}, V. 2010, \mnras, 401, 791

\bibitem[{{Springel} {et~al.}(2018){Springel}, {Pakmor}, {Pillepich}, {Weinberger}, {Nelson}, {Hernquist}, {Vogelsberger}, {Genel}, {Torrey}, {Marinacci}, \& {Naiman}}]{Springel2018_TNG}
{Springel}, V., {Pakmor}, R., {Pillepich}, A., {et~al.} 2018, \mnras, 475, 676

\bibitem[{{Ting} {et~al.}(2015){Ting}, {Conroy}, \& {Goodman}}]{Ting2015}
{Ting}, Y.-S., {Conroy}, C., \& {Goodman}, A. 2015, \apj, 807, 104

\bibitem[{{Tinsley}(1979)}]{Tinsley1979}
{Tinsley}, B.~M. 1979, \apj, 229, 1046

\bibitem[{{Vincenzo} \& {Kobayashi}(2018)}]{Vincenzo2018}
{Vincenzo}, F. \& {Kobayashi}, C. 2018, \mnras, 478, 155

\bibitem[{{Wang} \& {Lilly}(2023)}]{WangLilly2023}
{Wang}, E. \& {Lilly}, S.~J. 2023, \apj, 955, 55

\bibitem[{{Wang} {et~al.}(2024){Wang}, {Carrillo}, {Ness}, \& {Buck}}]{Wang2023}
{Wang}, K., {Carrillo}, A., {Ness}, M.~K., \& {Buck}, T. 2024, \mnras, 527, 321

\bibitem[{{Weinberger} {et~al.}(2018){Weinberger}, {Springel}, {Pakmor}, {Nelson}, {Genel}, {Pillepich}, {Vogelsberger}, {Marinacci}, {Naiman}, {Torrey}, \& {Hernquist}}]{2018MNRAS.479.4056W}
{Weinberger}, R., {Springel}, V., {Pakmor}, R., {et~al.} 2018, \mnras, 479, 4056

\bibitem[{{Xiang} {et~al.}(2019){Xiang}, {Ting}, {Rix}, {Sandford}, {Buder}, {Lind}, {Liu}, {Shi}, \& {Zhang}}]{Xiang2019_lamost}
{Xiang}, M., {Ting}, Y.-S., {Rix}, H.-W., {et~al.} 2019, \apjs, 245, 34

\end{thebibliography}
\bibliographystyle{aa}
\newpage

\begin{appendix}

\section{Galaxies with low correlation between [Fe/H] and \mbox{$\rm \text{R}_\text{birth}$}}\label{sec:appendix_lowCor}

\begin{figure*}[h!]
    \centering
     \includegraphics[width=17cm]{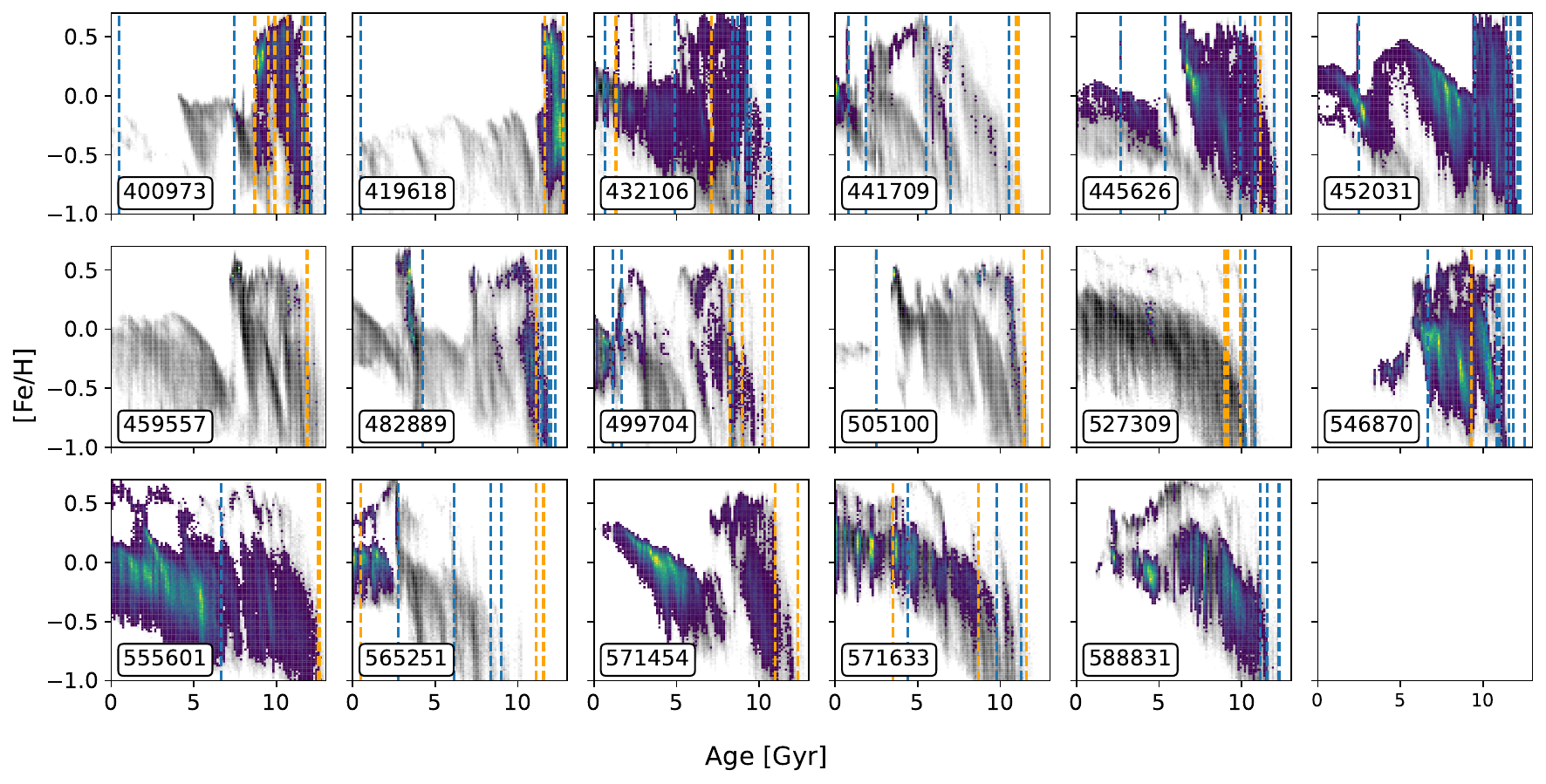}
\caption{Age--metallicity relation of the 17 galaxies that were removed from analysis in Section \ref{sec:results_gradFeh} due to the lack of linear metallicity gradient. The grey density map corresponds to the density of stellar particles that are currently bound and were born bound to the host galaxy. The colored density illustrates the density of our stellar disk sample ($R < 15$ kpc, $\mbox{$\rm \text{R}_\text{birth}$} < 15$ kpc, $|\mbox{$\rm \text{z}_\text{birth}$}| < 1$ kpc, |z| $< 1$ kpc, eccentricity $< 0.5$). The dashed lines show the merger events with a gas mass ratio $>25\%$ (blue) and $10-25\%$ (orange). Most of these galaxies are non-Milky Way-like due to too few disk stellar particles. The others are non-Milky Way-like due to a large spike in [Fe/H] in the inner kpc (e.g. subhalo 555601) causing a non-linear metallicity gradient (see also Figure \ref{fig:feh0est}).}
\label{fig:noCorPlot}
\end{figure*}

In Section \ref{sec:grad_Rbcorr} we showed that while most galaxies in the Milky Way and Andromeda analogues catalog of TNG50 \citep{Pillepich2023_MW_Sample} displayed a linear metallicity gradient, there were 17 galaxies that did not have a strong correlation between [Fe/H] and birth radius at a given age. Here, we discuss these galaxies in more detail to illustrate why they do not have a linear gradient, and how they also are non-Milky Way-like. Figure \ref{fig:noCorPlot} shows the age--metallicity relation of these 17 galaxies with their subhalo ID number given in the bottom left corner of each panel. The grey colormap corresponds to the density of stellar particles that are currently bound and were born bound to the host galaxy. The colored density corresponds to the density of our stellar disk sample ($R < 15$ kpc, $\mbox{$\rm \text{R}_\text{birth}$} < 15$ kpc, $|\mbox{$\rm \text{z}_\text{birth}$}| < 1$ kpc, |z| $< 1$ kpc, eccentricity $< 0.5$). The dashed lines correspond to merger events with a gas mass ratio $>25\%$ (blue) and $10-25\%$ (orange). For 10 of the 17 galaxies, there are minimal stellar disk particles, which span only a few Gyr. 6 of the galaxies (432106, 445626, 452031, 555601, 571633, 588831) exhibit a nonlinear gradient due to a large spike of [Fe/H] in the inner kpc (see Figure \ref{fig:feh0est} for a similar, less severe example). Of the 17 galaxies with poor [Fe/H]--\mbox{$\rm \text{R}_\text{birth}$} correlation, Subhlao 571454 appears to be the only one with a non-linear gradient due to a large dispersion in [Fe/H] for a given birth radius at a given time.

\FloatBarrier

\section{Extra Figures}

\begin{figure*}
    \centering
     \includegraphics[width=.95\textwidth]{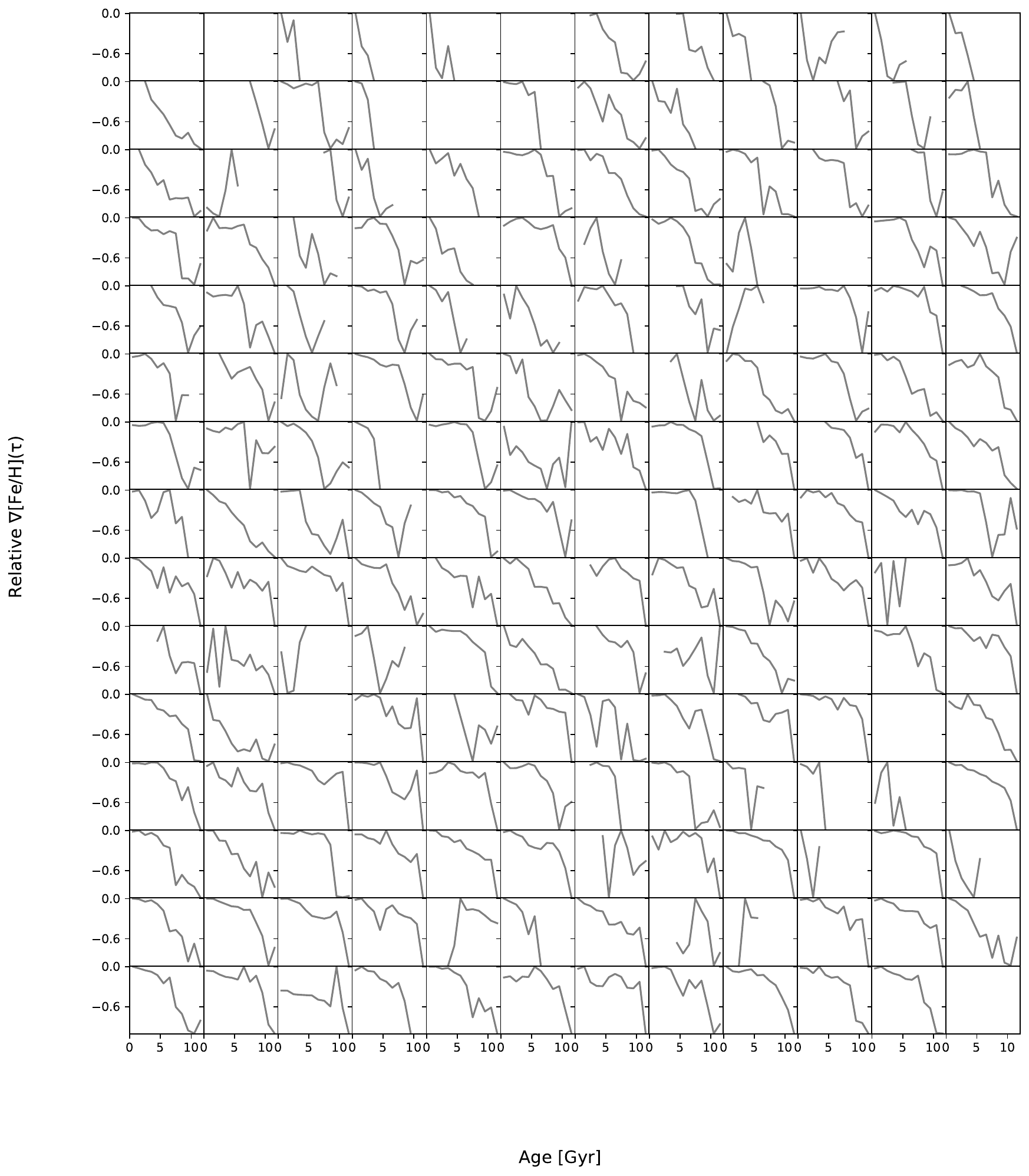}
\caption{Evolution of \mbox{$\rm \nabla [Fe/H](\tau)$} over time for each galaxy with a linear gradient ([Fe/H]--\mbox{$\rm \text{R}_\text{birth}$} correlation below -0.85) for at least the most recent 4 Gyr of star formation. We find that nearly all galaxies show a weakening in their [Fe/H] gradients over time. About 20 galaxies show a steepening in their metallicity gradient at older ages before flattening, consistent with what is now predicted for the Milky Way.}
\label{fig:bigGradPlot}
\end{figure*}

Figure \ref{fig:bigGradPlot}shows the relative metallicity gradient for each galaxy as a function of age. We find $\sim10-15$ galaxies show an initial steepening in their gradient before weakening over time, similar as found in the Milky Way (\citealt{Lu2022_Rb, Ratcliffe2023_enrichment}) and in external galaxies.

\begin{figure*}
    \centering
     \includegraphics[width=.95\textwidth]{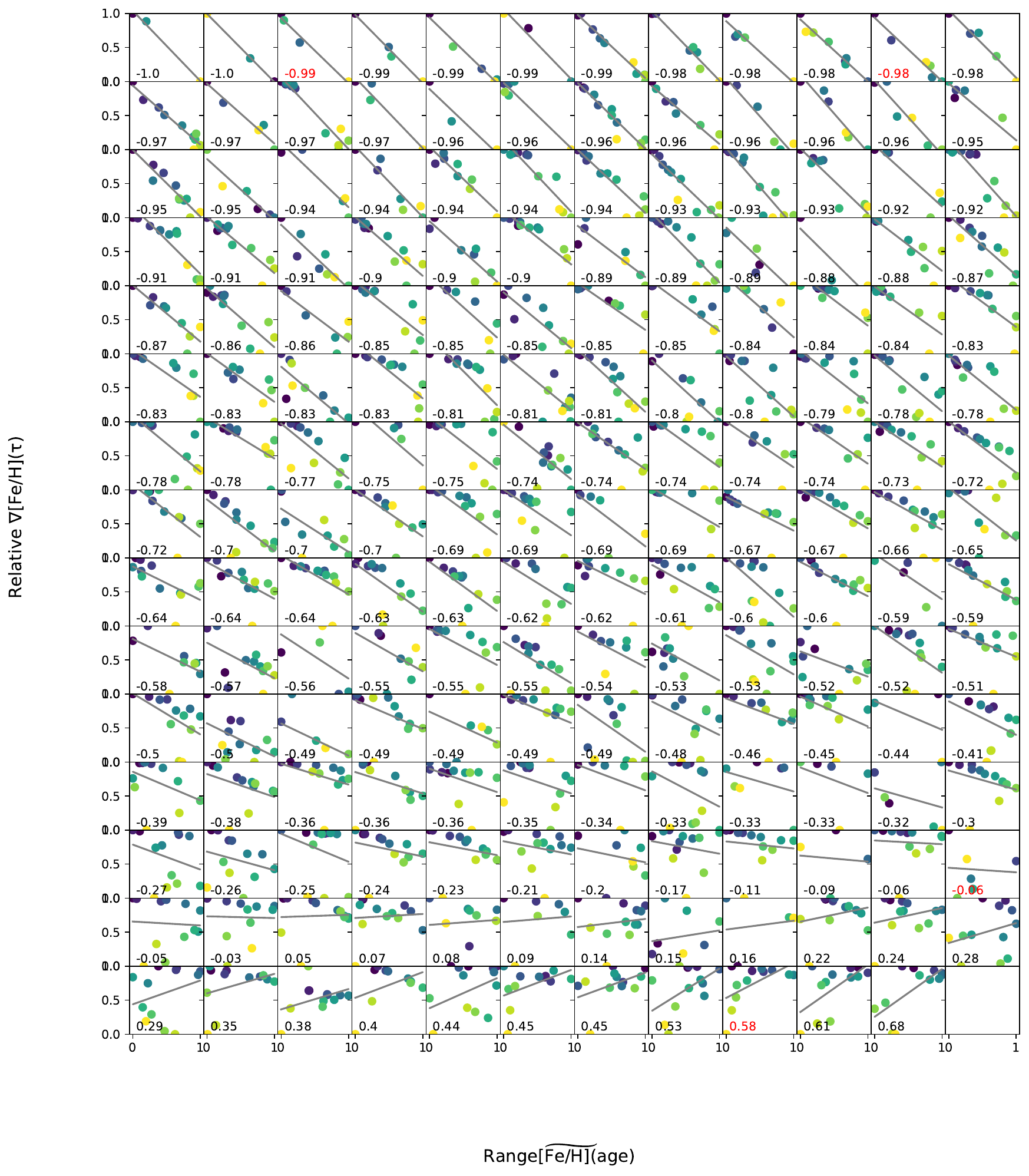}
\caption{Relationship between scatter in [Fe/H] and [Fe/H] radial birth gradient. Each point represents a different age bin, colored from youngest (blue) to oldest (yellow). The grey line is the line of best-fit. The correlation between \mbox{$\rm \nabla [Fe/H](\tau)$} and \rm \mbox{$\text{Range}\widetilde{\mbox{$\rm [Fe/H]$}}(age)$} is given in the bottom left of each panel. The red font indicates the galaxies analysed in Figure \ref{fig:outliers}.}
\label{fig:gradScatt_bigFig}
\end{figure*}

Figure \ref{fig:gradScatt_bigFig} shows the relative metallicity gradient versus the range in metallicity as a function of time. The correlation values are used in Section \ref{sec:grad_Scatcorr}. Many galaxies have a lower correlation value due to one or two outlier points (e.g. the first panel of row 11).

\begin{figure*}
    \centering
     \includegraphics[width=.95\textwidth]{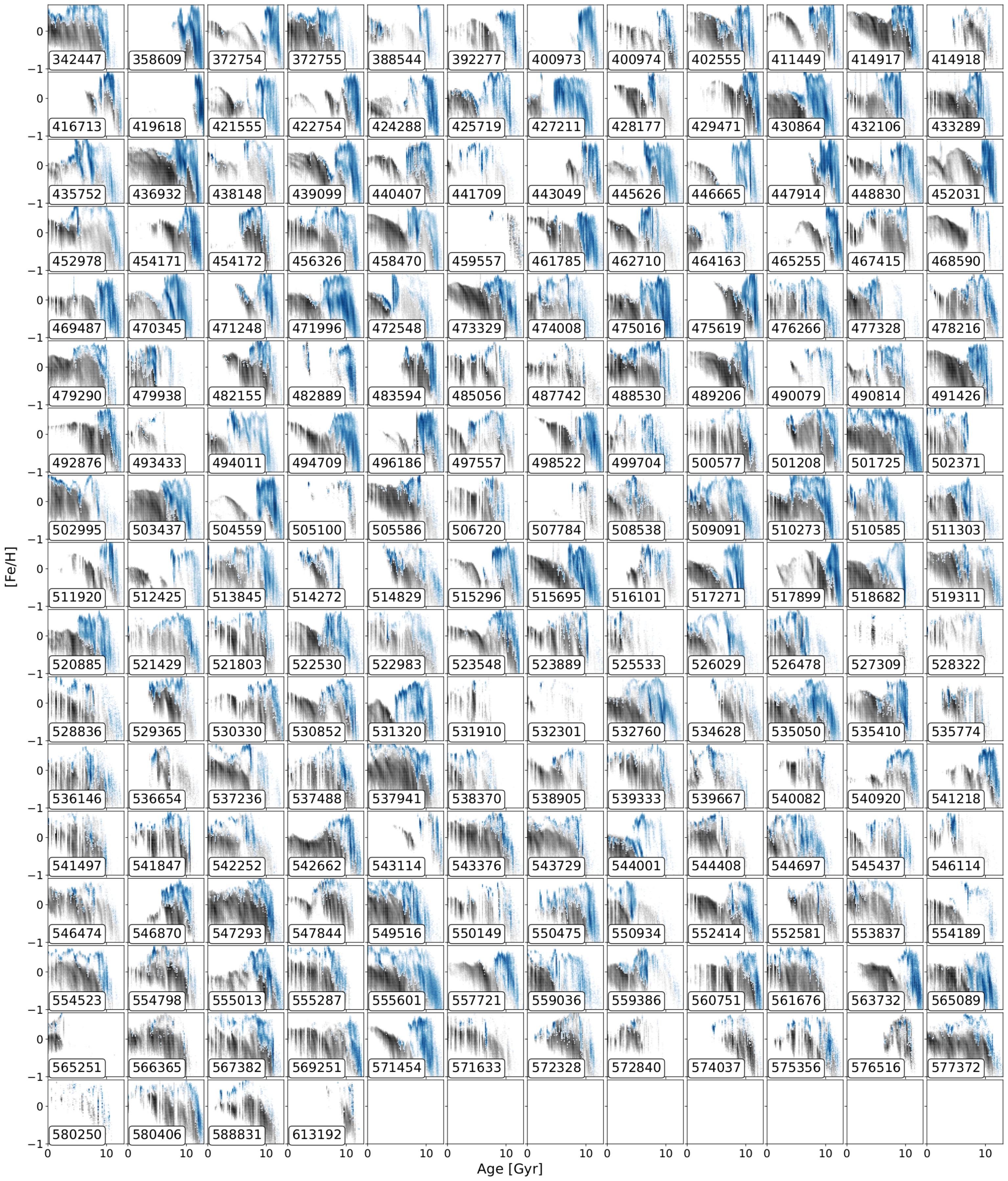}
\caption{Age--metallicity relation of disk stellar particles for each galaxy in our sample. The grey density refers to the disk sample, while the blue density corresponds to the density of the disk sample with $\text{R}_\text{birth} < 1$ kpc. Most galaxies show a halt in their star formation in their central regions.}
\label{fig:amr_bigFig}
\end{figure*}

\begin{figure*}
    \centering
     \includegraphics[width=.95\textwidth]{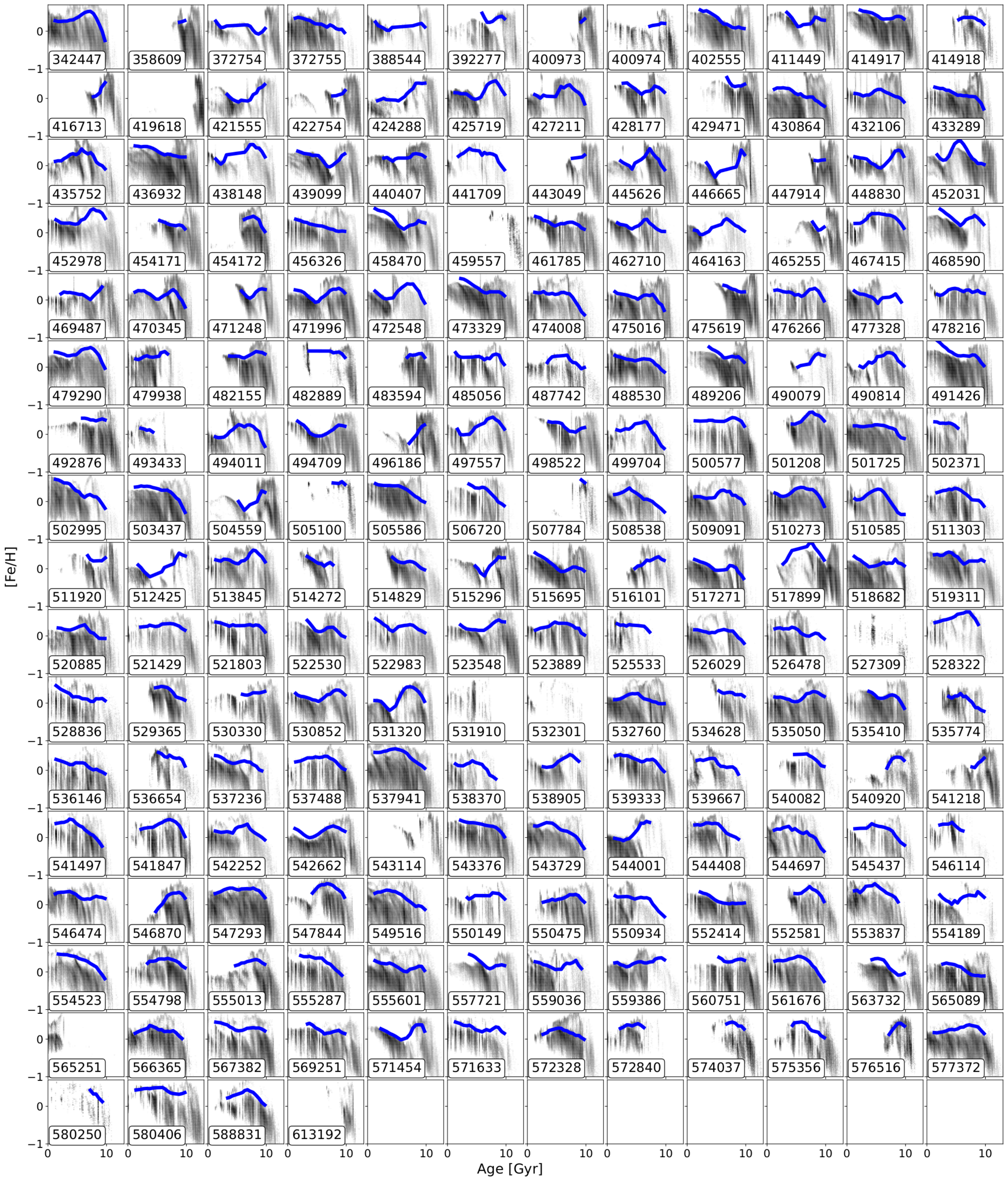}
\caption{Density of the age--metallicity relation for the disk stellar sample of each galaxy. The dark blue line corresponds to the projected central metallicity that is used as the anchor point to determine the metallicity profile for the genuine disk (see Section \ref{sec:results_feh0}). About half the sample shows a dilution in the projected central metallicity, while the other half shows a predominantly [Fe/H]-enriching center. If no blue line is shown, then there were not enough stars to determine the projection.}
\label{fig:feh0Proj_bigFig}
\end{figure*}

Figure \ref{fig:amr_bigFig} illustrates the age--metallicity relation for stellar disk particles (grey density map) and stellar disk particles with R$_\text{birth} < 1$ kpc (blue density) in each galaxy. Many of these galaxies show star formation quenched in their central kpc before redshift 0. Figure \ref{fig:feh0Proj_bigFig} also shows the age--metallicity relation for stellar disk particles in each galaxy, along with the projected central metallicty evolution (blue line). The projected central metallicity is the anchor point for the rest of the disk metallicity profile (Equation \ref{eqn:feh_Rb}) and is shown to be more representative of the disk than the inner few kpc (Section \ref{sec:results_feh0}). 

\begin{figure*}
     \centering
     \includegraphics[width=\textwidth]{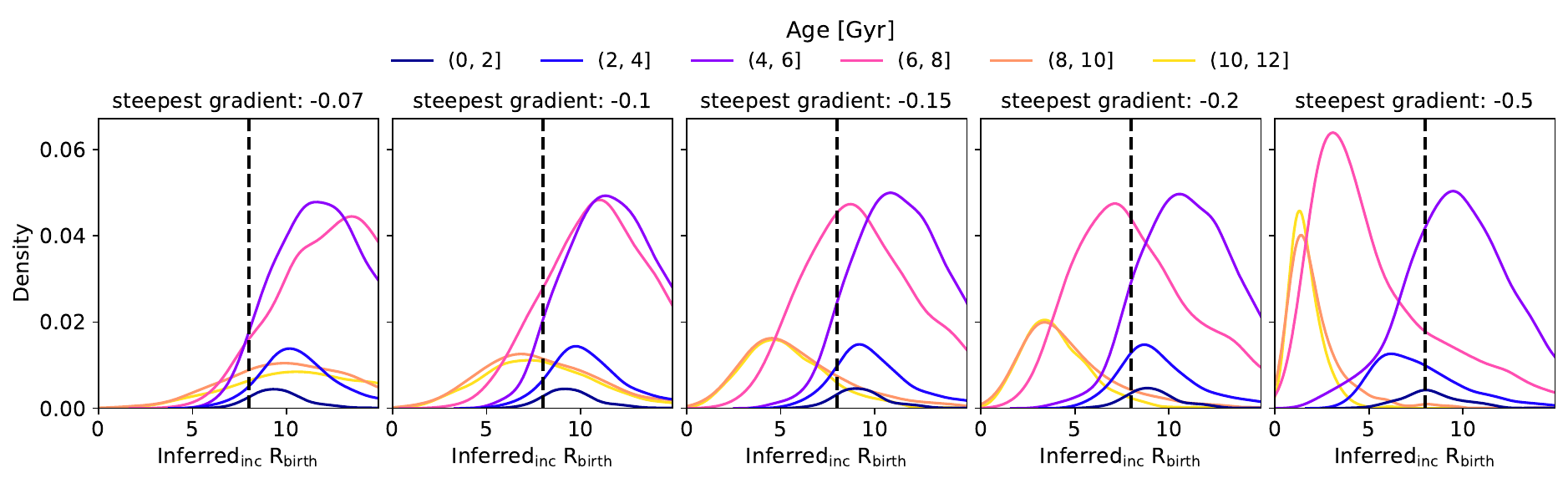}\\
\caption{Assuming different gradient strengths (in dex/kpc) changes the shape of the mono-age \mbox{$\rm \text{R}_\text{birth}$} populations for disk stellar particles currently near 8 kpc in subhalo 560751. However, varying the maximum strength of the gradient is never able to recover the correct trends as found in the second from the left panel of Figure \ref{fig:Rb_est}.} 
\label{fig:changeMaxGrad}
\end{figure*}

Figure \ref{fig:changeMaxGrad} illustrates how changing the value of the steepest gradient in recovering \mbox{$\rm \nabla [Fe/H](\tau)$} from \rm \mbox{$\text{Range}\widetilde{\mbox{$\rm [Fe/H]$}}(age)$} affects the \mbox{$\rm \text{R}_\text{birth}$} distributions of mono-age populations in the solar neighborhood. This example is using the incorrect red line from Figure \ref{fig:Rb_est} as the central metallicity evolution, and shows how even changing the value of the steepest gradient does not recover the proper distributions as given in the second panel of Figure \ref{fig:Rb_est}.

\end{appendix}

\end{document}